\documentclass[a4paper,onecolumn,11pt]{quantumarticle}
\pdfoutput=1
\usepackage[utf8]{inputenc}
\usepackage[T1]{fontenc}
\usepackage[english]{babel}
\usepackage{amsmath,amssymb,amsthm,mathtools,booktabs,array,graphicx,microtype,needspace,wrapfig,longtable}
\usepackage[numbers,sort&compress]{natbib}
\usepackage{hyperref}
\usepackage{tikz}
\usetikzlibrary{arrows.meta,positioning,fit}
\hypersetup{colorlinks=true,linkcolor=blue,urlcolor=blue,citecolor=blue,pdftitle={QEDist: Practical Exact Distance Certification for Stabilizer Codes},pdfauthor={Jong Yeon Lee}}
\newtheorem{proposition}{Proposition}

\newtheorem{theorem}[proposition]{Theorem}
\newcommand{\F}{\mathbb F_2}
\newcommand{\supp}{\operatorname{supp}}
\newcommand{\wt}{\operatorname{wt}}
\newcommand{\row}{\operatorname{row}}
\newcommand{\rank}{\operatorname{rank}}
\newcommand{\codepkg}{\textsf{QEDist}}
\begin{document}
\raggedbottom
\setlength{\emergencystretch}{1em}
\title{QEDist: Practical Exact Distance Certification for Stabilizer Codes}
\author{Jong Yeon Lee}
\email{jongyeon@illinois.edu}
\affiliation{Department of Physics, Illinois Quantum Information Science and Technology Center, The Grainger College of Engineering, University of Illinois at Urbana-Champaign, Urbana, Illinois 61801, USA}
\affiliation{Korea Institute for Advanced Study, Seoul 02455, South Korea}
\date{}
\begin{abstract}
We present \href{https://github.com/yeon-lee/QEDist}{\codepkg}, a method and software package that certifies the distance of qubit stabilizer codes.
Building on tree-search methods, \codepkg{} constructs candidate logical operators one qubit at a time, using sparse parity checks, stabilizer equivalence, and verified symmetries to reduce the work.
For a cutoff $W$, a completed search either finds a nontrivial logical operator of weight at most $W$ or proves that none exists.
Because satisfying one check can violate another, we use both satisfied and unsatisfied checks to lower-bound the weight needed to complete a logical operator, and prune branches whose bound exceeds the remaining budget, which substantially reduces net search time.
We benchmark \codepkg{} mainly on Calderbank-Shor-Steane (CSS) codes; the same formulation extends to non-CSS codes and to the fault distance of syndrome-extraction circuits.
For lifted-product (LP) codes, we further exploit the product structure to replace part of the full-code search with a smaller search on a classical constituent.
We establish exact distances for codes with up to several thousand qubits, including Abelian and non-Abelian LP codes whose distances were previously reported only as upper bounds.
In several seconds of search on one laptop core, \codepkg{} certifies exact distances up to 20 for many of these codes and fault distance 10 for the syndrome-extraction circuit of the $[[144,12,12]]$ code.
With orders-of-magnitude speedups over the tested exact solvers, \codepkg{} makes exact distance certification a practical tool for selecting and optimizing finite-length quantum codes.

\end{abstract}
\maketitle

\setcounter{tocdepth}{1}
\begingroup
\small
\hypersetup{linkcolor=quantumviolet}
\makeatletter
\patchcmd{\l@section}{1.0em}{0.4em}{}{}
\makeatother
\tableofcontents
\endgroup
\section{Introduction}

\begin{figure*}[!t]
\centering
\includegraphics[width=\textwidth]{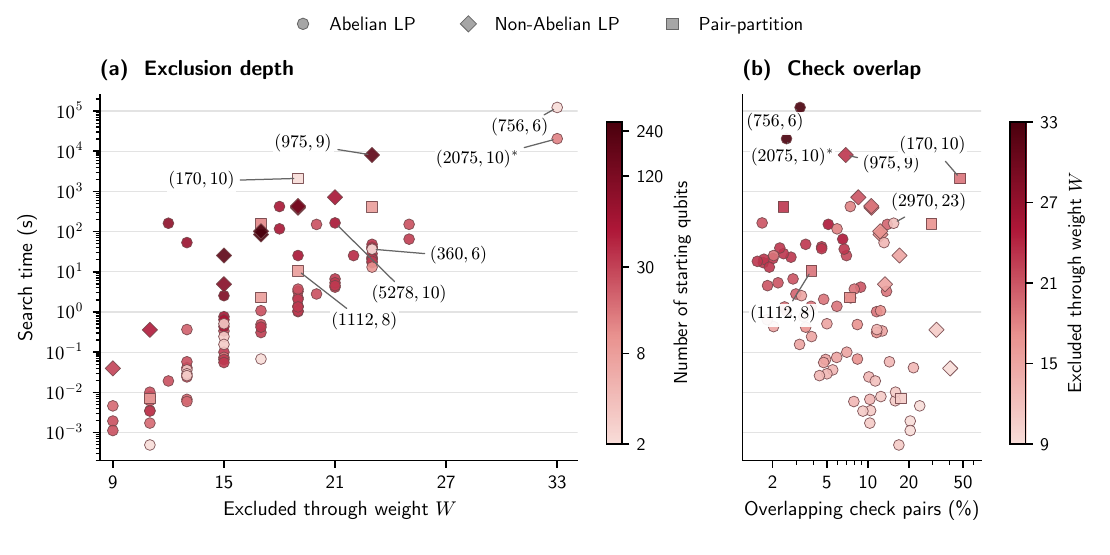}
\caption{Search time for completed exclusions versus (a) cutoff $W$ and (b) the fraction of overlapping check pairs.
Marker shapes identify code families, with bicycle and tricycle codes included under Abelian LP.
Color gives (a) the number of starting coordinates used per sector and (b) the exclusion cutoff $W$.
Labels identify selected cases by $(n,w)$, where $n$ is the number of physical qubits and $w$ is the maximum check weight.
Cutoffs and code parameters vary across points.
If all zero-syndrome supports in a sector have even weight, exclusion through an even cutoff also excludes the next odd weight.
All plotted runs use the weighted-check bound.
They use general CSS search except for $(2075,10)^{*}$, which uses LP-specific starting-point ordering and constituent reduction (Sec.~\ref{sec:lp-reduction}).
For Abelian and non-Abelian LP codes, these steps give $2.33$--$4.25\times$ speedups in separate benchmarks (Sec.~\ref{sec:lp-benchmark}).
Code distances, total search times, and input selection are given in Appendix~\ref{app:literature-catalogue}.
Search times are summed over all workers used to complete each exclusion
and rescaled to an Apple M3 Pro reference
(Sec.~\ref{sec:reduced-time}). 
}
\label{fig:catalogue-effort}
\end{figure*}

The distance is the smallest weight of an undetectable error that acts nontrivially on the encoded information.
For many quantum low-density parity-check (LDPC) codes, a reported distance upper bound leaves open the possibility of a substantially lighter logical operator.
Such an operator would weaken the protection suggested by the reported parameters and could change which finite-length code one chooses.
Determining the distance requires an explicit logical operator, called a witness, and a proof that no lighter logical operator exists.
Proving the lower bound is difficult because the number of possible supports is enormous.

For hypergraph-product codes built from full-row-rank classical check matrices, the distance is determined by the classical constituents~\cite{TillichZemor2014}, so computing it reduces to the smaller classical problems.
Lifted-product (LP) codes offer greater flexibility and can achieve larger distances, but their distance is not determined by the classical constituents~\cite{PanteleevKalachev2021,PanteleevKalachev2022}.
For LP codes, one has therefore often relied on randomized searches that supply logical witnesses and hence upper bounds on the distance~\cite{PryadkoShabashovKozin2022}.
In some cases, algebraic analyses constrain the low-weight logical operators that can occur~\cite{RaveendranDeclercqVasic2025}, and constituent and quotient searches give further upper bounds~\cite{Lee2026Spectroscopy,BhardwajEtAl2026Processors}.

For lower bounds, one can instead try to exclude every nontrivial logical operator of weight at most $W$, which establishes $d\ge W+1$.
Enumerating all supports up to this weight quickly becomes infeasible, but sparse checks strongly constrain how a candidate support can grow.
Exhaustive tree searches exploiting this structure were developed for stopping and trapping sets in classical LDPC codes~\cite{WangKulkarniPoor2009}.
Related enumeration algorithms support low-weight codeword searches and minimum-distance certification~\cite{RosnesYtrehus2009,RosnesEtAl2012}, and have been applied to CSS-code distances~\cite{MostadEtAl2025Bicycle}.
For quantum codes, Dumer, Kovalev, and Pryadko developed linked-cluster methods~\cite{KovalevDumerPryadko2013}; their irreducible-cluster algorithm adds qubits to satisfy violated parity checks, then tests whether the resulting undetectable operator acts nontrivially on the encoded information~\cite{DumerKovalevPryadko2015,DumerKovalevPryadko2016}.
In earlier studies of group-valued codes, including LP codes and pair-partition (PP) codes, we developed additional pruning rules within this tree search to accelerate code search~\cite{LeeEtAl2026DesignPrinciples,Lee2026RegularFrames}.

In this paper, we present \codepkg{} (Quantum Exact Distance) for exhaustive distance certification of qubit stabilizer codes. 
For many published LP codes whose distances were previously known only through upper bounds, \codepkg{} certifies the exact distance in seconds of search time on a laptop.
The search uses verified symmetries to reduce the required starting points and stabilizer equivalence to discard supports that cannot represent a minimum-weight logical operator.
We accelerate this search with a lower bound on the number of qubits that must be added to a partial support to satisfy all parity checks.
Checks that are already satisfied also matter, since adding a qubit to satisfy one check can violate another and force further additions.
We combine these requirements into a \emph{weighted-check bound}, accounting for overlaps between check supports.
When this bound exceeds the remaining weight budget, the branch is rejected before those additions are explored.
The general search uses the physical check matrices and the full stabilizer space, so it applies across construction families.
For suitable LP presentations, we additionally exploit their product structure to reduce the problem.

\begin{table*}[!t]
\centering
\small
\begin{tabular}{@{}lrrr@{}}
\toprule
Source family & Listed &  Exactly verified & Unresolved \\
\midrule
Monomial LP benchmarks~\cite{PanteleevKalachev2022,RaveendranEtAl2022GKP,XuEtAl2024AtomArrays,WebsterJacobHiggott2026Distance} & 18 & 18 & 0 \\
Logical spectroscopy~\cite{Lee2026Spectroscopy} & 13 & 6 & 7 \\
High-rate monomial LP codes~\cite{CainEtAl2026Shor,Lee2026Spectroscopy} & 3 & 3 & 0 \\
Radial codes~\cite{ScrubyHillmannRoffe2026Radial} & 10 & 10 & 0 \\
Weight-reduction inputs~\cite{SaboEtAl2024WeightReduction} & 3 & 3 & 0 \\
Bias-tailoring CSS input~\cite{RoffeEtAl2023BiasTailored} & 1 & 1 & 0 \\
Polynomial-domain LP codes~\cite{NouroziEtAl2026Polynomial} & 11 & 11 & 0 \\
Clustered-cyclic LP codes~\cite{GuEtAl2026QGPU} & 6 & 6 & 0 \\
Erasure-decoding LP inputs~\cite{KuoOuyang2026Erasure} & 5 & 5 & 0 \\
Finite-length distance examples~\cite{RaveendranDeclercqVasic2025} & 2 & 2 & 0 \\
Other Abelian matrix LP examples~\cite{OldRisplerMueller2024LCS,BhardwajEtAl2026Processors} & 2 & 2 & 0 \\
Cyclic spatially coupled example~\cite{LiuMarquardt2026} & 1 & 1 & 0 \\
Trivariate tricycle codes~\cite{JacobMcLauchlanBrowne2025Tricycle} & 6 & 6 & 0 \\
Bivariate bicycle codes~\cite{BravyiEtAl2024Bicycle} & 3 & 3 & 0 \\
Non-Abelian LP codes~\cite{BhardwajEtAl2026Processors,Hong2026NonAbelianLP} & 11 & 11 & 0 \\
\midrule
Total & 95 & 88 & 7 \\
\bottomrule
\end{tabular}
\caption{Distance status for the tabulated LP instances with $d>5$.
Of the $88$ exact distances verified here, $64$ match upper bounds reported in the cited sources.
Repeated presentations are counted once using the criteria in Appendix~\ref{app:survey-sources}, with the three reused Cain inputs assigned to their original family.
Appendix~\ref{app:literature-catalogue} lists all instances.}
\label{tab:survey-family-summary}
\end{table*}

\subsection{Main results}
\label{sec:main-results}

We apply \codepkg{} to a broad range of quantum LDPC codes, including Abelian and non-Abelian lifted-product constructions and pair-partition codes.
Table~\ref{tab:survey-family-summary} summarizes our distance-certification results for the surveyed families.

In seven tests of individual CSS sectors across six codes, \codepkg{} completes the requested exclusions in $0.17$--$7.94$ seconds on the Apple M3 Pro reference scale, including preparation and startup (Sec.~\ref{sec:benchmarks}).
On the first five LP tests in Table~\ref{tab:four-configurations}, the weighted-check bound reduces median search time by factors of approximately $3.1$--$21$ relative to the same search with the bound disabled, with all branching and other pruning rules fixed.
For LP inputs, the tensor-block structure guides the starting-point order and lets us replace part of the full-code search with a smaller constituent problem.
Together, these changes give a median speedup of $2.59$ over the general CSS search on the eleven tested monomial inputs with $d_Z\ge20$ (Sec.~\ref{sec:lp-benchmark}).

For the surveyed Abelian LP codes, we settle most previously reported upper bounds (Table~\ref{tab:survey-family-summary}).
For the $[[2075,87]]$ code of Nourozi et al.~\cite{NouroziEtAl2026Polynomial}, exclusion through weight $33$ and a weight-$34$ logical operator establish $d=34$.
We also establish $d=24$ for the $[[975,195]]$ non-Abelian LP code from the mitten family~\cite{BhardwajEtAl2026Processors}, using a verified central cyclic subgroup to reduce the search.
For the bivariate bicycle (BB) codes of Bravyi et al.~\cite{BravyiEtAl2024Bicycle}, we certify $[[360,12,24]]$ and $[[756,16,34]]$, reaching our deepest full-code exclusion through weight $33$.
The Abelian LP survey includes six trivariate tricycle (TT) codes with newly certified distances $14$ and $16$~\cite{JacobMcLauchlanBrowne2025Tricycle}.
We also certify the complementary $X$-sector distances of the $[[180,12,8]]$ and $[[432,12,12]]$ TT codes, establishing $(d_X,d_Z)=(20,8)$ and $(36,12)$, respectively (Table~\ref{tab:bb-tt-certification}).
These sector-resolved results quantify the asymmetric protection relevant to state preparation and logical operations.
Figure~\ref{fig:catalogue-effort} relates search effort to the completed exclusion cutoff and check overlap for both exact and unresolved cases.

For Bravyi et al.'s syndrome-extraction circuit for the $144$-qubit BB code~\cite{BravyiEtAl2024Bicycle}, we also confirm their conjectured fault distance of ten (Sec.~\ref{sec:circuit-applications}).

\section{The certification problem}
\label{sec:problem}
\subsection{Detecting nontrivial logical operators}
Up to an overall phase, an undetectable Pauli acts trivially on the encoded information exactly when it belongs to the stabilizer group.
To determine distance, we must therefore test both whether an operator has zero syndrome and whether it acts nontrivially.
For a CSS code, let $H_X$ and $H_Z$ be binary matrices with $n$ columns satisfying $H_XH_Z^{\mathsf T}=0$.
Their rows generate the $X$- and $Z$-type stabilizers, with $k=n-\rank H_X-\rank H_Z>0$ encoded qubits.
For the $Z$ sector, put $H=H_X$ and $S=\row H_Z$.
A binary support vector $z$ represents an undetectable operator when $Hz=0$, and a nontrivial logical operator when also $z\notin S$.
Thus
\begin{align}
d_Z=\min\{\wt(z):Hz=0,\ z\notin S\}.
\label{eq:sector-distance}
\end{align}
Interchanging $X$ and $Z$ gives $d_X$, and the CSS distance is $d=\min(d_X,d_Z)$.
All weights in this formulation count physical qubits.

We choose a cutoff $W$ and ask whether a nontrivial logical operator of weight at most $W$ exists.
A completed exclusion establishes $d_Z\ge W+1$.
Together with a checked logical operator of weight $W+1$, a completed exclusion through $W$ establishes $d_Z=W+1$.
Both sectors must reach the required lower bound to certify the full-code distance, and a witness in either sector supplies an upper bound.

\codepkg{} can test membership in $S$ directly by binary Gaussian elimination, even when the supplied generators are linearly dependent.
An equivalent test uses a precomputed logical detector matrix $L$, whose rows represent a complete logical basis of the opposite sector. A zero-syndrome candidate $z$ is logically nontrivial exactly when $Lz\ne0$.
The rows of $L$, together with those of $H$, span $S^\perp$, so
\begin{align}
\bigl(Hz=0,\ Lz=0\bigr)\quad\Longleftrightarrow\quad z\in S.
\label{eq:complete-detector}
\end{align}

\subsection{Physical inputs and symmetries}
\codepkg{} takes physical binary check matrices and optional symmetry generators given as permutations of the qubit indices.
LP constituents are expanded into these binary checks as described in Appendix~\ref{app:physical-inputs}.

A physical permutation symmetry preserves logical weight and can reduce the number of starting coordinates.
Most benchmarks use cyclic symmetries for this reduction.
Write the qubit coordinates as $(a,j)$, with block index $a$ and $j\in\mathbb Z/P\mathbb Z$.
The translation $\tau(a,j)=(a,j+1)$ shifts all blocks simultaneously.
\codepkg{} verifies that this translation preserves both check row spaces, allowing a translated generator to be a product of supplied generators.
The test therefore also covers redundant generating sets.
For non-Abelian LP codes, a central cyclic subgroup can supply such an action after reordering the coordinates (Appendix~\ref{sec:nonabelian-inputs}).
The relevant period is the order of the verified action.
For BB and TT inputs, the joint orbits of two or three verified independent translations leave two or three starting points, respectively (Table~\ref{tab:bb-tt-certification}).

\section{Exhaustive search with sparse checks}
\label{sec:search}
\begin{figure*}[t]
\centering
\includegraphics[width=\textwidth]{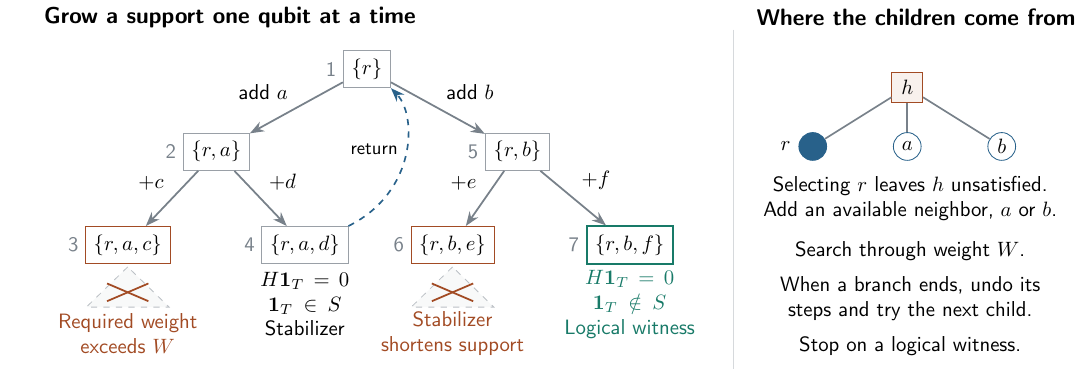}
\caption{Depth-first search from one starting coordinate.
Each node records a selected support $T$, and each edge adds one available neighbor of an unsatisfied check.
Numbers give the visit order in this schematic traversal.
One branch is pruned because satisfying the remaining checks would take the support above the cutoff. In the other, a stabilizer shortens the support.
The dashed arrow denotes return to an ancestor before exploring its next child.
}
\label{fig:tree-search}
\end{figure*}

A check is \emph{unsatisfied} if it overlaps the selected support in an odd number of qubits and \emph{satisfied} if the overlap is even. An undetectable support satisfies every check.
The search adds a qubit from an unsatisfied check and explores the resulting branches one at a time, returning to the next unused choice when a branch ends (Fig.~\ref{fig:tree-search}).

\subsection{Starting points and branching}
A verified cyclic translation reduces the required starting coordinates from $n$ to $n/P$~\cite{ZhangSiegel2011,Lee2026RegularFrames}.
For any nonempty support, take its first occupied block and translate one occupied coordinate in that block to position zero.
It therefore suffices to start once at $(a,0)$ for each block $a$, excluding all earlier blocks [Fig.~\ref{fig:algorithm-overview}(a)].
This covers every orbit, including orbits shorter than $P$, and may cover an orbit more than once.
With several verified generators, the same rule uses the smallest-index coordinate in each joint orbit and excludes earlier coordinates.

\Needspace{5\baselineskip}
A nonempty undetectable support is \emph{support-irreducible} if it contains no smaller nonempty undetectable support~\cite{DumerKovalevPryadko2015,DumerKovalevPryadko2016}.
Every shortest logical has this property, since a smaller undetectable support would either be a lighter logical or a stabilizer whose removal lowers its weight.

At a node, let $T$ be the selected nonempty support, $s=H\mathbf 1_T$ its syndrome, and $F$ the forbidden coordinates.
Initially, $F$ contains the coordinates earlier than the starting point.
Selected coordinates are also unavailable.
For a zero-syndrome target support $U$ containing $T$ and disjoint from $F$, the additional qubits $U\setminus T$ form a \emph{completion} of $T$.
For each check $c$, its available neighborhood is
\begin{align}
N_c(T,F)=\supp(H_c)\setminus(T\cup F).
\label{eq:available}
\end{align}
\Needspace{5\baselineskip}
If $s\ne0$, choose an unsatisfied check $c$. Every completion must include an element of $N_c$.
Order its available neighbors as $q_1,\ldots,q_m$.
Child $i$ includes $q_i$ and forbids $q_1,\ldots,q_{i-1}$.
Each possible completion is assigned to exactly one child, determined by its first neighbor in this order~\cite{LeeEtAl2026DesignPrinciples} [Fig.~\ref{fig:algorithm-overview}(b)].
These starting-point and sibling exclusions are the only rules that add coordinates to $F$.

To limit the number of children, we choose an unsatisfied check with the fewest available neighbors.
When $s=0$, we test membership in $S$ and end the branch.
Any extension would contain the current undetectable support and hence could not be a shortest logical.

\begingroup
\setlength{\textwidth}{.98\textwidth}
\begin{figure*}[t]
\centering
\includegraphics[width=\textwidth]{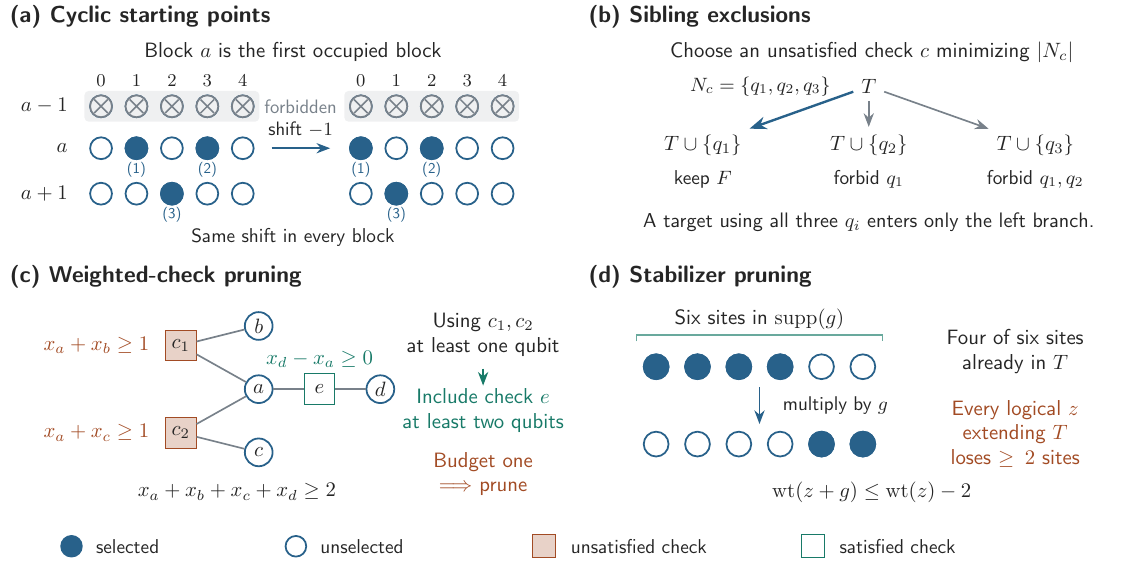}
\caption{Four mechanisms that reduce the search for a shortest logical.
(a) Each row is a block of $P$ qubits. Shifting all blocks by the same amount moves an occupied site of the first occupied block $a$ to position zero. Earlier blocks are forbidden (crossed-out row).
(b) Ordered children inherit the selected support $T$ and forbidden set $F$. The highlighted branch retains a target using all three neighbors.
(c) Adding $a$ satisfies both unsatisfied checks $c_1,c_2$ but violates the satisfied check $e$, forcing the addition of $d$. Avoiding $a$ requires both $b$ and $c$. Including $e$ raises the lower bound on additional weight from one to two. The variables $x_v$ indicate qubit additions.
(d) The selected support $T$ overlaps a weight-six stabilizer $g$ in four sites. Every logical support $z$ containing $T$ therefore overlaps $g$ in at least four sites. Multiplying by $g$ gives an equivalent logical operator $z+g$ at least two qubits lighter, so this branch cannot contain a shortest logical.}
\label{fig:algorithm-overview}
\end{figure*}

\endgroup

\subsection{Bounding the weight needed for completion}
A branch can be rejected when satisfying all checks would require more qubits than the cutoff permits.
Before the search starts, we check whether parity reduces the cutoff.
If the all-ones vector belongs to $\row H$, every undetectable operator has even weight, so an odd cutoff $W$ can be replaced by $W-1$.
A list of check rows summing to the all-ones vector certifies this reduction.
Write $W_{\mathrm{eff}}=W-1$ in this case and $W_{\mathrm{eff}}=W$ otherwise.
Appendix~\ref{app:correctness} gives the proof and the corresponding criterion for binary monomial LP constituents.

Let $b=W_{\mathrm{eff}}-|T|$ be the number of further qubits permitted.
An unsatisfied check with no available neighbor rules out completion immediately.
Each added qubit touches at most $w_c$ checks, where $w_c$ is the largest column weight of $H$.
If $\wt(s)>bw_c$, the remaining $b$ qubits cannot touch every unsatisfied check, so we prune the branch.

The locations of the unsatisfied checks can require more additions than their number alone suggests.
If $p$ unsatisfied checks have pairwise disjoint $N_c$, each requires a different new qubit, so
\begin{align}
\text{additional weight}\ \ge p,
\qquad p>b\ \Longrightarrow\ \text{prune}.
\label{eq:packing}
\end{align}
Such a collection is called a packing, and any packing gives a valid lower bound.
This disjoint-neighborhood bound is used in our construction study~\cite{LeeEtAl2026DesignPrinciples} and has an intersection-graph formulation in classical hitting-set solvers~\cite{BlasiusEtAl2021HittingSet}.
Earlier classical LDPC searches also used a linear-programming bound on the additional weight needed to cover unsatisfied checks~\cite{ZhangSiegel2011}.

\subsection{Using satisfied checks}
Adding a qubit to satisfy one check can violate another and force further additions.
Suppose the unsatisfied checks $c_1,c_2$ have available neighborhoods $\{a,b\}$ and $\{a,c\}$, and a satisfied check $e$ has neighborhood $\{a,d\}$ [Fig.~\ref{fig:algorithm-overview}(c)].
The first two checks share $a$, so their disjoint-neighborhood bound is one.
Adding $a$ satisfies both but violates $e$, forcing the addition of $d$.
Avoiding $a$ requires both $b$ and $c$.
In either case, at least two additions are needed.

We can express this argument using $x_q\in\{0,1\}$ to indicate whether qubit $q$ is added.
Checks $c_1$ and $c_2$ require $x_a+x_b\ge1$ and $x_a+x_c\ge1$, and check $e$ requires $x_d-x_a\ge0$.
Adding these inequalities gives $x_a+x_b+x_c+x_d\ge2$.
The inequality from $e$ reduces the coefficient of $x_a$ from two to one and introduces $x_d$, so each possible addition contributes at most one to the sum.
An already satisfied check can therefore rule out a completion that the disjoint-neighborhood bound allows.

Replacing the exact parity constraints by these weaker inequalities gives a parity relaxation~\cite{FeldmanWainwrightKarger2005}.
Related linear-programming relaxations provide classical minimum-distance bounds~\cite{BurshteinGoldenberg2010} and support exact distance computation by branch-and-bound or branch-and-cut~\cite{KehaDuman2010,HelmlingEtAl2014}.
\codepkg{} combines these inequalities with nonnegative integer weights, keeping the coefficient of each available qubit at most two, so a combined requirement above $2b$ rules out completion within $b$ additions.
Since the bound is evaluated at many search nodes, we limit the number of integer updates to control its total evaluation cost (Appendix~\ref{app:weighted-checks}).

The same inequality can test a proposed addition by fixing that qubit's contribution and reducing the remaining budget by one.
If the bound shows that completing this choice would exceed the remaining weight budget, we skip its branch.

\subsection{Using stabilizer equivalence}
Multiplying a logical operator by a stabilizer preserves its encoded action and can reduce its weight.
For every $g\in S$, a minimum-weight logical $z$ satisfies
\begin{align}
|\supp(z)\cap\supp(g)|\le\frac{\wt(g)}2,
\label{eq:coset}
\end{align}
since otherwise $z+g$ is a lighter representative of the same nonzero logical class~\cite{Lee2026RegularFrames} [Fig.~\ref{fig:algorithm-overview}(d)].
Because the selected support only grows, the violation persists throughout the subtree.

\codepkg{} updates these overlaps as each qubit is added or removed, using the supplied stabilizer generators, and prunes as soon as it detects a violation.

\Needspace{9\baselineskip}
\subsection{Completed exclusions}
\label{sec:software}

\begin{theorem}[Sound cutoff exclusion]
\label{thm:search}
Let $S$ be a subspace of $\ker H$ over $\F$, and let any supplied permutation generators preserve both spaces.
Apply the starting-point, branching, stopping, and pruning rules above at cutoff $W$.
If every required branch finishes without finding a word in $\ker H\setminus S$, then $d_Z\ge W+1$.
\end{theorem}
The proof moves a minimum-weight logical operator to a required starting point, then follows choices that add qubits from its support.
The branching and pruning rules preserve a path to that operator.
The full argument is given in Appendix~\ref{app:correctness}.

A verified permutation exchanging the CSS check spaces can transfer a sector result to the other sector.
Otherwise, both sectors are searched.
\codepkg{} checks the inputs, symmetries, and witnesses, then verifies that the completion records cover all required work with consistent settings (Appendix~\ref{app:workflow-records}).
Verifying the exhaustive exclusion from these records requires replaying the search.

In validation runs, \codepkg{} reproduces previously certified distances, including all fourteen high-check-weight controls, and agrees with the two completed calculations by the exact Boolean optimizer MaxCDCL (Appendices~\ref{app:conditional-followup} and~\ref{app:final-protocol}).

\subsection{Using lifted-product structure}
\label{sec:lp-reduction}
An LP presentation provides a further reduction by separating the two tensor blocks of physical qubits.
For constituents $A$ and $B$ of sizes $r_A\times s_A$ and $r_B\times s_B$ over a cyclic group algebra, the $Z$-sector check matrix is the binary expansion of
\begin{align}
H_X=\left[A\otimes I_{s_B}\quad I_{r_A}\otimes B^\dagger\right],
\label{eq:lp-block-check}
\end{align}
where $\dagger$ combines transpose with inversion of group elements.
Write a physical support as $(u,v)$, with $u$ arranged as an $s_A\times s_B$ array of group-algebra entries.
We place all second-block orbits first and search from one starting point in each.
A minimum-weight logical with $v\ne0$ can be translated to include the starting point in its first occupied second-block orbit.
It is therefore compatible with the exclusions of earlier orbits, while all first-block qubits remain available.
We skip full-code searches from first-block starting points and handle $v=0$ through the constituent reduction below.

For a zero-syndrome operator with $v=0$, each of the $s_B$ columns separately satisfies the binary check equations of $A$.
Let $K$ be the kernel of the binary expansion of $A$.
Let $E_j a$ denote the placement of a constituent vector $a$ in column $j$ of the first block, with zeros on all other qubits.
Every zero-syndrome operator $(u,0)$ therefore decomposes as $\sum_j E_j a_j$ with $a_j\in K$.
If the sum is logical, at least one of these zero-syndrome terms is logical, since a sum of stabilizers is a stabilizer.
That term has no greater weight, because the columns have disjoint physical supports.

A column word may be a stabilizer in one placement and logical in another.
With $S=\row H_Z$, define the common trivial subspace
\begin{align}
S_{\cap}=\{a\in K\,|\,E_j a\in S\text{ for every }j\}.
\label{eq:lp-common-trivial}
\end{align}
For $a\in K$, we have $a\notin S_{\cap}$ if and only if at least one placement $E_j a$ is logical, with weight $\wt(a)$.
We therefore run one constituent search for a word in $K\setminus S_{\cap}$, then check its physical placements.
Translation within each column preserves $K$ and $S_{\cap}$, so the same cyclic starting-point reduction applies.
Theorem~\ref{thm:search} applies to this smaller problem as well.
If the second-block search and the constituent search both complete without a witness through cutoff $W$, then $d_Z\ge W+1$.

The implementation checks directly on the physical matrices that setting $v=0$ leaves independent column equations with a common kernel.
For the $X$ sector, the same argument uses rows of the first block and the constituent $B$.
Although derived for cyclic group algebras, these physical-matrix conditions also hold for the non-Abelian $[[975,195,24]]$ mitten code, allowing the same reduction to accelerate its certification (Sec.~\ref{sec:lp-benchmark}).

\section{Performance and the sources of computational gain}
\label{sec:benchmarks}

\subsection{Inputs and timing}
\label{sec:reduced-time}
We compare \codepkg{} with other exact methods on six CSS codes, measuring the time required to exclude logical operators through the weight cutoffs listed in Table~\ref{tab:final-comparison}.
The seven tests include both $X$ and $Z$ sectors for the $1428$-qubit code and one sector for each remaining code.

m4riCC is the connected-cluster search in \texttt{dist-m4ri}~v0.9.0, which grows candidate logical supports through unsatisfied parity checks~\cite{DumerKovalevPryadko2016,DistM4RI}.
\codepkg{} uses the same support-growth principle with the check-selection and pruning rules developed in Sec.~\ref{sec:search}.

MaxCDCL is an exact solver for Boolean optimization problems that learns constraints from conflicts encountered during the search~\cite{LiEtAl2022MaxCDCL}.
We encode parity checks and nontrivial logical action as Boolean constraints, then use it to minimize operator weight or exclude operators through a specified weight cutoff~\cite{ChenJafariLai2026}.

Every method receives the same physical input and known witness, and verified parity reductions are applied to all methods.
The connected-cluster comparison also uses the same cyclic starting points as \codepkg{} (Appendix~\ref{app:final-protocol}).
Search time includes bound evaluation. Unless stated otherwise, total elapsed time also includes preparation and startup.
All reported runtimes and time limits use an Apple M3 Pro reference scale (Appendix~\ref{app:timing-conventions}).

\begin{table}[!b]
\centering\small
\setlength{\tabcolsep}{5pt}
\begin{tabular}{@{}rrrcccrrr@{}}
\toprule
& & & & & & \multicolumn{3}{c}{Total elapsed time (s)}\\
\cmidrule(l){7-9}
$n$\hphantom{~[00]} & $k$ & $P$ & Sector & $W$ & Row / column wt. & \codepkg{} & m4riCC & MaxCDCL\\
\midrule
320~\cite{Lee2026RegularFrames} & 64 & 16 & $Z$ & 9 & 9 / 3--6 & 0.169 & 2.62 & 12.1\\
476~\cite{Lee2026RegularFrames} & 56 & 7 & $Z$ & 13 & 8--10 / 3--6 & 0.446 & T & 340\\
714~\cite{WebsterJacobHiggott2026Distance} & 100 & 21 & $Z$ & 15 & 8 / 3--5 & 0.323 & T & T\\
1122~\cite{ZhouEtAl2026Surgery} & 148 & 33 & $Z$ & 19 & 8 / 3--5 & 2.43 & T & T\\
1428~\cite{LiuMarquardt2026} & 186 & 42 & $X$ & 17 & 8 / 3--5 & 0.589 & T & T\\
1428~\cite{LiuMarquardt2026} & 186 & 42 & $Z$ & 17 & 8 / 3--5 & 0.728 & T & T\\
1768~\cite{WebsterJacobHiggott2026Distance} & 224 & 52 & $Z$ & 21 & 8 / 3--5 & 7.94 & T & T\\
\midrule
\multicolumn{6}{@{}l}{Completed sector tests} & $7/7$ & $1/7$ & $2/7$\\
\bottomrule
\end{tabular}
\caption{Seven CSS-sector exclusions through weight $W$. $P$ is the verified cyclic period. Row and column weights are for the binary syndrome matrix of the tested sector.
For five codes, a verified symmetry exchanges the $X$ and $Z$ sectors, so one search suffices. For the $1428$-qubit code, the $X$ and $Z$ check spaces have different ranks ($622$ and $620$), so we search both sectors.
Times include preprocessing and use the Apple M3 Pro reference scale. T denotes the approximately $570$-second limit. m4riCC uses the repaired \texttt{dist-m4ri}~v0.9.0 build (Appendices~\ref{app:timing-conventions} and~\ref{app:final-protocol}).}
\label{tab:final-comparison}
\end{table}

\subsection{How the search work is reduced}
\label{sec:weighted-ablation}
\begin{figure}[!t]
\centering
\includegraphics[width=.97\linewidth]{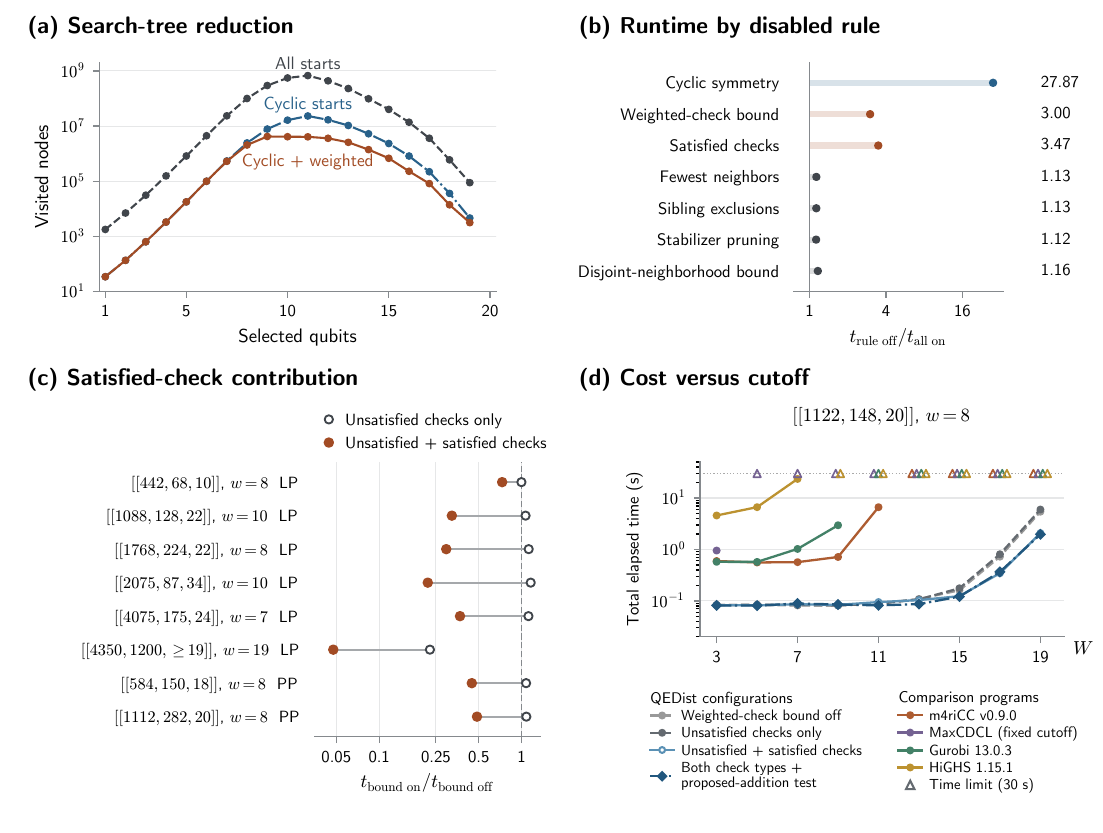}
\caption{Search reduction and runtime.
(a) Visited nodes for the $Z$-sector exclusion through weight $21$ in the $[[1768,224,22]]$ code, using all $1768$ starting coordinates, $34$ cyclic starts, and cyclic starts with the weighted-check bound.
(b) Effect of disabling each rule, relative to the all-rules median of $7.41$ seconds.
Disabling the fewest-neighbors rule selects the first unsatisfied check.
Disabling ``Satisfied checks'' retains only the inequalities from unsatisfied checks.
(c) Ratios of median search times $t_{\mathrm{bound\ on}}/t_{\mathrm{bound\ off}}$ for eight $Z$-sector examples at the cutoffs in Table~\ref{tab:four-configurations}.
The numerator uses the checks indicated in the legend, and the denominator disables the weighted-check bound.
Labels give $[[n,k,d]]$ and maximum check weight $w$, using certified lower bounds when $d$ is unresolved.
(d) Total elapsed time versus the requested cutoff for the $[[1122,148,20]]$ LP code.
For (d), we run MaxCDCL on a fixed-cutoff encoding (Appendix~\ref{app:mechanism-cutoffs}).
Timings are three-run medians, with ranges shown in (b).
Open triangles mark the $30$-second limit for unfinished searches.
Parity reduces each odd cutoff $W$ in (a,d) to $W-1$.
Times use the Apple M3 Pro reference scale, with (b,c) comparing search times and (d) total elapsed times (Appendices~\ref{app:final-protocol} and~\ref{app:mechanism-cutoffs}).}
\label{fig:search-profile}
\end{figure}

The weighted-check bound substantially reduces the search tree even after cyclic symmetry has reduced the starting points.
For the $Z$-sector exclusion through weight $21$ in the $[[1768,224,22]]$ LP code~\cite{WebsterJacobHiggott2026Distance}, enabling the bound reduces the number of visited nodes from $86.3$ million to $23.6$ million, chiefly by pruning deeper branches (Fig.~\ref{fig:search-profile}(a)).

Across fourteen LP and PP sector problems, the bound using only unsatisfied checks speeds up the higher-check-weight examples but slows several others.
Adding satisfied-check inequalities reduces median search time by factors of $1.37$--$5.28$ relative to that bound, with branching and all other pruning rules fixed (Fig.~\ref{fig:search-profile}(c) and Table~\ref{tab:four-configurations}).
With both types of check, all fourteen exclusions finish faster than with the weighted-check bound disabled.
For example, the $1112$-qubit PP input takes $10.6$ seconds with the bound disabled, $11.4$ seconds using only unsatisfied checks, and $5.14$ seconds including both types of check.

Runtime gains also depend on the cost of each test.
Testing proposed additions reduces the number of visited nodes further but changes runtime only slightly at these cutoffs (Appendix~\ref{app:mechanism-cutoffs}).
In the $1768$-qubit example, the inexpensive disjoint-neighborhood bound saves time by rejecting some branches before the weighted-check bound is evaluated, with little change in the number of visited nodes (Fig.~\ref{fig:search-profile}(b)).

\subsection{Comparison with other exact methods}
\label{sec:final-comparison}
\codepkg{} completes all seven sector exclusions in Table~\ref{tab:final-comparison}.
With a budget of approximately $570$ seconds, m4riCC completes only the $320$-qubit case.
MaxCDCL completes the $320$- and $476$-qubit cases by proving the supplied witnesses optimal and reaches the time limit on the other five sectors.

In the single-worker comparison on two polynomial LP codes~\cite{NouroziEtAl2026Polynomial}, \codepkg{} certifies distance $20$ for the $600$-qubit code and excludes logical operators through weight $27$ for the $2075$-qubit code (Appendix~\ref{app:polynomial-comparison}).
Under the common $300$-second search limit, HiGHS completes neither exclusion and Gurobi reaches the limit on the $600$-qubit input.

\subsection{Effort as the cutoff increases}
\label{sec:cutoff-sweep}
Raising the cutoff allows heavier supports and can substantially increase the work.
We measure this growth by varying the exclusion cutoff for the $425$- and $1122$-qubit LP inputs, with the latter shown in Fig.~\ref{fig:search-profile}(d).
For the $1122$-qubit input, all four \codepkg{} configurations complete the deepest tested exclusion, through weight $19$.
Table~\ref{tab:shared-cutoffs} compares median elapsed times for \codepkg{} and each solver at the highest tested cutoff completed by both methods.
Across these comparisons, the other solvers took $11.5$ to $273$ times as long as \codepkg{}.

At shallow cutoffs, input preparation and startup account for most of \codepkg{}'s elapsed time.
As the cutoff increases, search takes a larger share of the total time and the bound has a greater effect.
For the $1122$-qubit input at $W=19$, the weighted-check bound reduces median elapsed time from $5.42$ to $2.08$ seconds.

\subsection{Additional gain from LP structure}
\label{sec:lp-benchmark}
We compare the LP reduction of Sec.~\ref{sec:lp-reduction} with the general CSS search on the nineteen monomial LP inputs in the Webster et al.\ benchmark~\cite{WebsterJacobHiggott2026Distance}.
Both configurations use the same cyclic symmetry, cutoff, and internal branching and pruning rules.
For these $3\times5$ constituents, the reduction replaces $34$ full-code starting points with nine in the second block and five in the smaller constituent problem.

Both configurations complete all nineteen $Z$-sector exclusions.
For the eleven cases with $d_Z\ge20$, the reordered starting points and constituent reduction together give a median speedup of $2.59$, with individual factors $2.33$--$4.25$ (Table~\ref{tab:lp-specialization}).
Across these eleven exclusions, the total node count falls from $580.4$ million to $158.0$ million.

Changing only the starting-point order gives a median speedup of $2.06$ relative to the same CSS baseline.
This ordering changes which coordinates are excluded from later searches, and therefore which branches the pruning rules can eliminate.
Relative to that reordered CSS search, the constituent reduction gives a further median speedup of $1.27$, with individual factors $1.21$--$1.87$.
For six inputs whose CSS searches finish in less than $0.01$ seconds, preparation overhead makes the LP configuration slower.

The reduction also gives a $3.12$-fold speedup for the non-Abelian $[[975,195,24]]$ mitten code, lowering total search time from $8056$ to $2579$ seconds on the reference scale (Table~\ref{tab:mitten-certification}).
The runs plotted in Fig.~\ref{fig:catalogue-effort} use general CSS search, with the LP reduction for $[[2075,87,34]]$.

\begin{table}[tbp]
\centering\small
\begin{tabular}{@{}rrrrrrr@{}}
\toprule
$n$\hphantom{~[00]} & $k$ & $d_Z$ & $W_{\mathrm{eff}}$ & $t_{\mathrm{CSS}}$ (s) & $t_{\mathrm{LP}}$ (s) & $t_{\mathrm{CSS}}/t_{\mathrm{LP}}$\\
\midrule
1054~\cite{PanteleevKalachev2022}\hphantom{0} & 140 & 20 & 18 & 2.18 & 0.51 & 4.25\\
1428~\cite{XuEtAl2024AtomArrays} & 184 & 24 & 22 & 52.56 & 13.36 & 3.94\\
1768~\cite{WebsterJacobHiggott2026Distance} & 224 & 22 & 20 & 6.44 & 2.18 & 2.96\\
1972~\cite{WebsterJacobHiggott2026Distance} & 248 & 20 & 18 & 1.03 & 0.42 & 2.44\\
2040~\cite{WebsterJacobHiggott2026Distance} & 256 & 20 & 18 & 1.29 & 0.50 & 2.59\\
2074~\cite{WebsterJacobHiggott2026Distance} & 260 & 24 & 22 & 27.42 & 9.93 & 2.76\\
2278~\cite{WebsterJacobHiggott2026Distance} & 284 & 22 & 20 & 5.18 & 1.91 & 2.71\\
2482~\cite{WebsterJacobHiggott2026Distance} & 308 & 24 & 22 & 20.67 & 8.22 & 2.51\\
2720~\cite{WebsterJacobHiggott2026Distance} & 336 & 22 & 20 & 4.36 & 1.82 & 2.39\\
2890~\cite{WebsterJacobHiggott2026Distance} & 356 & 24 & 22 & 25.15 & 9.91 & 2.54\\
3230~\cite{WebsterJacobHiggott2026Distance} & 396 & 24 & 22 & 19.35 & 8.31 & 2.33\\
\bottomrule
\end{tabular}
\caption{Combined gain from starting-point ordering and constituent reduction on the eleven monomial inputs with $d_Z\ge20$ from Ref.~\cite{WebsterJacobHiggott2026Distance}.
Times are two-run medians on the Apple M3 Pro reference scale, including LP-specific preparation and both component searches.
Complete logical detectors are prepared before timing in this experiment (Appendix~\ref{app:lp-benchmark}).
The completed exclusion through $W_{\mathrm{eff}}$, together with even-weight parity and a checked witness, establishes $d_Z$.
The $(1428,184)$ input here differs from the $(1428,186)$ input in Table~\ref{tab:final-comparison}.}
\label{tab:lp-specialization}
\end{table}

\section{Distance certification across code families}
\label{sec:applications}
Among the surveyed Abelian LP codes with $d>5$, we establish that $58$ of the $65$ reported distance upper bounds are exact.
These include all $17$ upper bounds in the monomial-LP benchmark family (Table~\ref{tab:survey-family-summary}).
Using the LP reduction, we certify $d=34$ for the $[[2075,87]]$ code by excluding logical operators through weight $33$ in $5.7$ worker-hours of search and verifying a weight-$34$ witness (Appendix~\ref{sec:survey-case-details}).
The general CSS search also certifies the radial code $[[736,18,26]]$~\cite{ScrubyHillmannRoffe2026Radial}, with $d_X=26$ and $d_Z\ge26$ (Table~\ref{tab:survey-all}).

Certifying the code distance can require less work than determining both sector distances exactly.
For the $[[1428,186]]$ code, $d_X=18$ and $19\le d_Z\le28$ already establish $d=18$.

Separate certification of $d_X$ and $d_Z$ is particularly useful for asymmetric codes, where the overall distance $d=\min(d_X,d_Z)$ reports only the weaker sector's protection.
The TT sector distances in Sec.~\ref{sec:main-results} illustrate how separate $X$- and $Z$-sector certificates could inform magic-state preparation and cultivation~\cite{MenonEtAl2026MagicTricycles,GidneyShuttyJones2024Cultivation}.
In our logical CCZ-state cultivation study, \codepkg{} enabled exact distance evaluation and efficient screening of candidate lifted bicycle codes~\cite{HirasakiLee2026CCZCultivation}.

Blocklength alone is a poor guide to the cost of certification.
We certify distance $16$ for the $[[5220,1674]]$ spectroscopy example, while several smaller codes in the same study remain unresolved~\cite{Lee2026Spectroscopy}.
Figure~\ref{fig:catalogue-effort}(b) plots the fraction of pairs of supplied checks in the same CSS sector that share at least one qubit.
At comparable total search times, larger overlap fractions generally accompany smaller exclusion cutoffs, although the code parameters and search settings also vary.

The $[[2970,1188]]$ code illustrates why different pruning rules are needed.
Its supplied stabilizer generators have weight $22$ or $23$, so a partial support of weight at most $11$ cannot be shortened by multiplying by any single supplied generator.
The weighted-check bound can still exclude branches when satisfying all checks would exceed the weight budget.

For Abelian LP codes with higher check weights, searches using the proposed-addition test improve several lower bounds and certify $[[1025,45,26]]$ and $[[5278,1480,22]]$~\cite{NouroziEtAl2026Polynomial,CainEtAl2026Shor,Lee2026Spectroscopy} (Appendix~\ref{app:conditional-followup}).
For the additional scalar-by-matrix LP examples of Panteleev and Kalachev~\cite{PanteleevKalachev2021}, we establish $d=24$ for the $[[882,24]]$ code, previously bounded by $18\le d\le24$.
For their $[[1270,28]]$ code, we raise the lower bound from $d\ge16$ to $d\ge24$ (Table~\ref{tab:survey-secondary}).

We also revisit the PP codes whose distances were certified with an earlier form of this tree search~\cite{LeeEtAl2026DesignPrinciples}.
For the $[[170,8,20]]$ PP code (Table~\ref{tab:pp-certification}), a verified $C_{85}$ action leaves two starting coordinates, and sector exchange makes one sector sufficient.
The exclusion through weight $19$ reproduces $d=20$ using $35.1$ worker-minutes on the Apple M3 Pro reference scale (Fig.~\ref{fig:catalogue-effort}).

\subsection{Non-Abelian lifted products}
\label{sec:nonabelian-applications}
We show that the reported upper bounds are tight for the non-Abelian LP codes $[[630,126,20]]$, $[[780,156,22]]$, and $[[975,195,24]]$~\cite{BhardwajEtAl2026Processors}.
We also certify Hong's metacyclic examples as $[[390,78,16]]$, $[[550,110,18]]$, and $[[625,125,20]]$~\cite{Hong2026NonAbelianLP} (Table~\ref{tab:mitten-certification}).

For the $975$-qubit code, the verified central $C_5$ action leaves $195$ starting coordinates per sector in the general CSS search.
Both sectors are excluded through weight $23$, and a checked weight-$24$ $X$ logical gives $d_X=d=24$ and $d_Z\ge24$.
The LP reduction brings the total search time to $43.0$ worker-minutes on the reference Apple M3 Pro scale (Sec.~\ref{sec:lp-benchmark}).

\subsection{Circuit fault distance}
\label{sec:circuit-applications}
Clifford syndrome-extraction circuits with independent Pauli faults give the same search problem, with each faulty operation counted once~\cite{WebsterJacobHiggott2026Distance}.
A detector model records the detection events and logical observables flipped by each fault, and the search seeks a smallest fault set with zero detector flips and nontrivial logical action.
The search branches on an unsatisfied detector and applies the weighted-check bound to detector parities.
Replacing part of a fault set by fewer faults with identical detector and logical effects gives a pruning rule analogous to stabilizer multiplication.

For the $[[144,12,12]]$ gross code, \codepkg{} confirms the circuit fault distance of ten conjectured by Bravyi et al.\ for their single noisy syndrome-extraction cycle followed by two noiseless cycles~\cite{BravyiEtAl2024Bicycle}.
Using lattice translations and fault exchanges, the exclusion through nine faults takes $4.32$ and $3.41$ seconds of search time in the $X$ and $Z$ sectors, respectively, on one Apple M3 Pro core (Appendix~\ref{app:bb-circuit}).
In each sector, a checked set of ten faults produces a logical error while leaving every syndrome measurement unchanged.

\section{Discussion}
Exact distance certification makes comparisons of finite-length quantum codes more reliable by establishing which reported upper bounds are attained.
For several codes with thousands of qubits and distances $20$--$24$, certification takes seconds to tens of seconds on the Apple M3 Pro reference scale.
The measured speedups over the tested exact solvers make exact distance a practical criterion for selecting and optimizing candidate codes.

The formulation extends to non-CSS qubit stabilizer codes by encoding the three Pauli choices at each qubit as binary columns.
In this encoding, the smallest number of selected Pauli choices that forms a nontrivial logical operator equals the physical code distance, as shown in Appendix~\ref{app:noncss-encoding}.

The search can also be parallelized by assigning groups of subtrees (shards) to separate workers (Appendix~\ref{app:partitioned-execution}).
This can yield near-linear speedups in elapsed time when shard workloads are balanced and overhead is small.
For the $1768$-qubit benchmark, four workers give a median $3.31$-fold speedup over one worker (Appendix~\ref{app:calibration}).

Further improvements could adapt the effort spent evaluating bounds to the remaining weight allowance and local check structure.
Compact exclusion proofs would allow a small independent program to check the lower bounds.
Lean-QEC provides a related approach through formally verified reductions to satisfiability and machine-checked distance certificates~\cite{EhatammEtAl2026LeanQEC}.
A quantitative relation between physical check structure and search cost remains the central open problem for predicting which codes can be certified within a practical budget.

\section*{Acknowledgement}
The computations were performed locally and through the Illinois Campus Cluster Program at the University of Illinois, Urbana-Champaign.
J.Y.L. is supported by a faculty startup grant at the University of Illinois, Urbana-Champaign.
The author acknowledges the use of AI tools from OpenAI for brainstorming ideas, running numerical searches, making figures, and editing the text.
The author has reviewed all content and takes full responsibility for the results.

\section*{Data and software availability}
\label{sec:data-availability}
\codepkg{} is available under the MIT license at \url{https://github.com/yeon-lee/QEDist}, together with example inputs, selected logical witnesses, and verification tools.
The complete benchmark archive, including physical inputs, witnesses, and search records, is available from the author upon request.
The prototype underlying \codepkg{} was developed in late July~2026 and was subsequently used for distance certification in Refs.~\cite{LeeEtAl2026DesignPrinciples,Lee2026RegularFrames,HirasakiLee2026CCZCultivation}.

\appendix
\section{Correctness of the exclusion rules}
\label{app:correctness}
The proof combines the irreducible-support argument of Refs.~\cite{DumerKovalevPryadko2015,DumerKovalevPryadko2016} with the validity of the pruning rules.
Call a search node $(T,F)$ \emph{compatible} with a support $U$ if $T\subseteq U$ and $F\cap U=\varnothing$.

\begin{proof}[Proof of Theorem~\ref{thm:search}]
Suppose $d_Z\le W$, and let $z$ be a minimum-weight logical operator with support $U$.
If a nonzero $y\in\ker H$ were supported on a proper subset of $U$, then either $y\notin S$ is a lighter logical or $y\in S$ and $z+y$ is a lighter logical.
Hence $z$ is support-irreducible.

Order the verified symmetry orbits by their smallest coordinates, using singleton orbits when no symmetry is supplied.
A verified symmetry moves an occupied coordinate of the first occupied orbit to that orbit's smallest coordinate, preserving weight and logical nontriviality.
Relabel the transformed operator and its support as $z$ and $U$.
The node at this starting coordinate is then compatible with $U$.

At a compatible node, every unsatisfied check has an available neighbor in $U\setminus T$.
The child that adds the first such neighbor remains compatible, while every other child adds a coordinate outside $U$ or forbids that neighbor.
Since parity reduction lowers an odd cutoff only when every kernel word has even weight, $\wt(z)\le W_{\mathrm{eff}}$, so $U\setminus T$ is a completion within the remaining budget.
It therefore satisfies the syndrome-count and disjoint-neighborhood bounds of Sec.~\ref{sec:search} and the weighted-check and proposed-addition inequalities of Appendix~\ref{app:weighted-checks}.
A stabilizer overlapping $T$ in more than half its weight would shorten $z$, contradicting minimality.
The compatible path thus survives every pruning rule, and support irreducibility prevents it from stopping at a zero-syndrome proper subset of $U$.
It reaches $z$, which the exact stabilizer-membership test identifies as a witness, so a completed witness-free search implies $d_Z>W$.
\end{proof}

\paragraph*{Parity certificates.}
Parity preprocessing is certified by $\alpha^{\mathsf T}H=\mathbf 1^{\mathsf T}$.
Every $z\in\ker H$ then obeys $\wt(z)\equiv\alpha^{\mathsf T}Hz=0\pmod 2$.
For fully populated binary monomial LP constituents with dimensions as in Appendix~\ref{app:physical-inputs} and $r_A\equiv s_B\pmod 2$, choose an $r_A\times s_B$ binary matrix with odd row and column sums.
Such a matrix exists whenever its dimensions have the same parity.
Select the corresponding check-block rows of $H_X$ and sum all $P$ translates of each.
Every qubit block occurs an odd number of times, yielding the physical all-ones vector independently of the exponents and the parity of $P$.
The $H_Z$ statement follows using an $s_A\times r_B$ selection matrix.
Conversely, every $H_Z$ row lies in $\ker H_X$ and has weight $r_A+s_B$, so unequal parity supplies an odd-weight stabilizer.
The same argument applies to $H_X$ and $\ker H_Z$.

\subsection{Search procedure}
\label{app:search-order}
The recursion tracks $T$, its syndrome, and $F$, with cached check neighborhoods and stabilizer overlaps.
For the disjoint-neighborhood bound, \codepkg{} first builds a greedy packing, considering smaller neighborhoods first.
Near the cutoff, it can also search for $b+1$ mutually disjoint neighborhoods within a fixed effort limit, pruning only when it finds such a packing.
If parity preprocessing gives $W_{\mathrm{eff}}=0$, no positive-weight support remains to search.
Otherwise the search starts from the required singleton supports.
\begin{quote}
\footnotesize
\begin{verbatim}
Search(T, F):
  if resource limit: record incomplete; return
  if a tested stabilizer shortens T: return
  if syndrome(T) = 0:
    test T modulo stabilizers; return
  b = W_eff - |T|
  if b <= 0 or a completion bound exceeds b: return
  choose an unsatisfied check with fewest available neighbors
  if it has no available neighbor: return
  for q in its ordered available neighbors:
    if no completion bound excludes this addition:
      Search(T union {q}, F)
    F = F union {q}
    if witness or incomplete: return
\end{verbatim}
\end{quote}
Each call has its own forbidden set, and backtracking restores the parent's cached state.
Testing for zero syndrome before checking the remaining weight budget retains logical operators of weight exactly $W_{\mathrm{eff}}$.

\subsection{The weighted-check bound}
\label{app:weighted-checks}
Let $V$ be the available coordinates and $x_v\in\{0,1\}$ indicate additions to the selected support $T$.
An unsatisfied check requires at least one addition.
If an addition touches a satisfied check, at least one more addition must touch it to restore even parity.
Every completion therefore obeys the following inequalities on the available neighborhoods of Eq.~\eqref{eq:available}.
\begin{align}
\sum_{v\in N_c}x_v&\ge1 &&\text{if }s_c=1,\\
\sum_{u\in N_e\setminus\{v\}}x_u-x_v&\ge0 &&\text{if }s_e=0,\ v\in N_e.
\label{eq:residual-parity}
\end{align}
Combining these inequalities with nonnegative integer multipliers gives $\beta\le\sum_{v\in V}c_vx_v$, where $\beta$ is the sum of the multipliers for unsatisfied checks and the coefficients $c_v$ may have either sign.
For any $D\ge0$ such that every $c_v\le D$, a completion within budget $b$ must therefore satisfy
\begin{align}
\beta\le\sum_{v\in V}c_vx_v\le D\sum_{v\in V}x_v\le Db.
\label{eq:weighted-check-pruning}
\end{align}
Thus $\beta>Db$ excludes the node, using integer arithmetic throughout.
This argument applies to every kernel completion, including stabilizers, and hence preserves any shortest logical compatible with the node.
The optional test of proposed additions uses the same inequality to reject individual children without recomputing the bound.
Conditioning on the addition of coordinate $q$ gives $\beta-c_q\le D(b-1)$, so a violation rules out that child.
The parent coefficients must be retained across recursive calls, and the rejected coordinate still receives the usual sibling exclusion.

The implementation greedily increases the multipliers for unsatisfied checks, allowing a bounded number of satisfied-check corrections to keep every coefficient at most $D$.
Successful updates are retained, and failed attempts restore the previous coefficients.
The effort limits select which valid inequality is found, while Eq.~\eqref{eq:weighted-check-pruning} establishes its safety.
The source code specifies the exact update order and effort limits.

\subsection{Partitioned execution}
\label{app:partitioned-execution}
A deterministic branching rule defines a rooted tree for each required starting coordinate.
Splitting the first child list into disjoint ranges partitions its subtrees, provided a worker still applies the exclusions generated by earlier siblings outside its assigned range.

An alternative divides the work into fixed groups of subtrees (shards) by assigning each node at a chosen depth to one group, using a deterministic hash modulo the number of groups.
The nodes at this depth form the search frontier.
Each shard traverses the same prefix, evaluates the same pruning rules, and descends below the frontier only for its assigned nodes.
Nodes that terminate before the frontier are checked in the common prefix.
Each frontier node has exactly one owning shard.
Completion requires every shard for every starting coordinate to finish without a witness.

\subsection{Non-CSS stabilizer distance}
\label{app:noncss-encoding}
The same binary search can determine general Pauli distance by assigning three unit-cost choices $X_i,Y_i,Z_i$ to each physical qubit.
Let $H$ record their commutation parities with the stabilizer generators and $L$ their commutation parities with a complete set of $2k$ logical Pauli operators.
Minimizing $\wt(e)$ subject to $He=0$ and $Le\ne0$ gives the physical Pauli distance.
Every weight-$w$ Pauli has an encoding with $w$ selected choices.
Conversely, multiple choices at one qubit multiply, up to phase, to a single Pauli or the identity, preserving syndrome and logical action without increasing weight.
Taking $S=\ker H\cap\ker L$ gives exactly the quotient search of Eq.~\eqref{eq:sector-distance}.
The kernel $S$ includes the local relations among the three choices as well as encodings of physical stabilizers.
The exclusion theorem therefore applies with ordinary binary weight, whose minimum on $\ker H\setminus S$ equals the physical Pauli distance.

\section{Benchmark details}
\label{app:reproducibility}
\subsection{Physical inputs and verification records}
\label{app:physical-inputs}
For binary constituents, take $R=\F[t]/(t^P-1)$ and constituents $A\in R^{r_A\times s_A}$ and $B\in R^{r_B\times s_B}$.
With $\dagger$ denoting transpose and $t\mapsto t^{-1}$, we use the LP checks
\begin{align}
H_X&=\left[A\otimes I_{s_B}\quad I_{r_A}\otimes B^\dagger\right],\\
H_Z&=\left[I_{s_A}\otimes B\quad A^\dagger\otimes I_{r_B}\right].
\label{eq:lp-checks}
\end{align}
Replacing $t$ by the cyclic permutation matrix expands these into commuting binary checks of length $n=P(s_As_B+r_Ar_B)$.
For quaternary constituents over $\mathbb F_4$, expansion in a fixed self-dual trace basis gives binary length $n=2P(s_As_B+r_Ar_B)$.
We compute $k$ and test symmetries on the expanded binary checks.
The $[[1428,186]]$ input is represented over $C_{42}$ using the isomorphism $C_3\times C_{14}\cong C_{42}$~\cite{LiuMarquardt2026}.

\label{app:workflow-records}
Each run records the input and program by content hash, together with its sector, requested and effective cutoffs, and partition.
Input checks verify CSS commutation, binary ranks, and the supplied symmetry and parity certificates.
Every witness is checked for physical weight, zero syndrome, and nontrivial logical action.
A new lower bound requires completed searches from all assigned starting coordinates, covering every subtree.

\subsection{Cyclic reduction for non-Abelian inputs}
\label{sec:nonabelian-inputs}
Consider an LP presentation built from left and right regular matrices of a finite group $G$, with a central cyclic subgroup $C\le Z(G)$.
Choose a generator $c$ of $C$.
Multiplication by $c$ commutes with both regular actions, so its simultaneous action on the coordinate copies of $G$ preserves both CSS check spaces~\cite[Appendix~D.3]{Hong2026NonAbelianLP}.
Every orbit on each coordinate copy has length $P=|C|$.
Ordering the coordinates orbit by orbit puts this permutation into block-cyclic form and leaves $n/|C|$ starting coordinates.
The physical row-space tests are then applied to the reordered checks.
A reduction to $n/|G|$ would require verification of the larger physical symmetry.

\subsection{Timing and implementation conventions}
\label{app:timing-conventions}
Search time measures the compiled search routine, including bound evaluation.
Unless stated otherwise, total elapsed time includes input preparation, validation, program startup, initialization, and solving.
Compilation precedes timing.
The catalogue reports \emph{total search time}, which sums the time spent searching across workers and can exceed elapsed time when workers run in parallel.

The Apple M3 Pro defines the reference scale for all reported times and time limits.
Local timings are measured directly on this machine.
The controlled cluster comparisons and catalogue reruns use AMD EPYC~7713 processors.
Cluster runtimes and time limits are divided by $\kappa\simeq2.11$ for reporting, including each worker's contribution to mixed-host totals.
This factor is the geometric mean of ratios from three compiled-search workloads, each measured three times on both hosts.
The ratios range from $1.81$ to $2.37$.
Converted times are approximate, and the same factor is used for older cluster records without a recorded processor model.
For program comparisons, the same conversion is applied to baseline times.
Each controlled comparison uses the same host for its configurations, and uniform scaling preserves their timing ratios.

\codepkg{} uses \texttt{-O3} for the controlled comparisons and the circuit benchmark.
These benchmarks use GCC~11.5.0 on the cluster and Apple Clang~21.0.0 locally.
The LP tensor-block comparison adds \texttt{-DNDEBUG -march=native}.
The weighted-check bound uses $D=2$ and at most four satisfied-check corrections per update.

The baseline implementations are \texttt{dist-m4ri}~v0.9.0, MaxCDCL from the 2023 MaxSAT Evaluation (MSE~2023), Gurobi~13.0.3, and HiGHS~1.15.1.
On the cluster, m4riCC and MaxCDCL use the same GCC compiler with \texttt{-O3 -DNDEBUG}.

\subsection{Program and search-rule comparisons}
\label{app:final-protocol}
The $320$- and $476$-qubit controls come from Ref.~\cite{Lee2026RegularFrames}, the $1122$-qubit input from Ref.~\cite{ZhouEtAl2026Surgery}, and the other inputs from Refs.~\cite{WebsterJacobHiggott2026Distance,LiuMarquardt2026}.
\begin{sloppypar}
The seven-sector program comparison uses one cluster core per task.
Through support weight eight, ties between equally small available neighborhoods favor those touching more unsatisfied checks in total.
Neighbors are visited in decreasing order of the number of unsatisfied checks they touch, with ties resolved by coordinate index.
The comparison uses the fewest-neighbors rule with this tie-breaker, the disjoint-neighborhood and weighted-check bounds, and stabilizer pruning (Sec.~\ref{sec:search}).
The proposed-addition test is disabled.
\end{sloppypar}

Each baseline receives one run with a total elapsed-time limit of approximately $570$ seconds.
The \codepkg{} times are medians of two runs on the same core per sector, except for $1428$-$Z$, which has one run.
Table~\ref{tab:final-comparison} reports these values and single-run baseline times on the reference scale defined in Appendix~\ref{app:timing-conventions}.
The connected-cluster implementation receives the same cyclic starting coordinates, exclusions of earlier coordinate blocks, and verified parity reduction as \codepkg{}.
Its repeated launches and preprocessing count toward total elapsed time.
We patched the m4riCC coordinator so that assigning a starting coordinate and registering its worker happen atomically.
Without this patch, a run could finish before all starting coordinates were searched.

For six sectors in Table~\ref{tab:final-comparison}, we used MaxCDCL with default preprocessing to minimize logical weight, with a complete logical detector basis computed from the physical checks and a checked weight-$(W+1)$ witness as the initial upper bound.
The two completed optimizations gave the independently certified sector distances $10$ and $14$ for the $320$- and $476$-qubit codes, respectively.
For the $1428$-qubit $Z$ sector, the available witness has weight $28$. We tested $W=17$ directly with the cyclic encoding and variable elimination disabled (Appendix~\ref{app:mechanism-cutoffs}). The run reached this time limit.

The measurements in Fig.~\ref{fig:search-profile}(a,b) run on the cluster.
For (a), a build instrumented to count nodes at each depth reproduces the total node counts of the timed build across $64$ search partitions on $32$ cores, counting each shared starting node once.
For (b), three repetitions per configuration use $24$ single-core workers, each completing within the approximately $850$-second limit.
In (b), search times include bound evaluation, with depth counters and overlap tie-breaking disabled.

\subsection{Polynomial LP inputs and mixed-integer comparison}
\label{app:polynomial-comparison}
The two instances are the larger polynomial lifts chosen for decoder simulations by Nourozi et al.~\cite{NouroziEtAl2026Polynomial}.
The source reports $d\le20$ at lift $24$, obtained with Gurobi, and $d\le34$ at lift $83$.
For a binary $Z$ support $x$ and cutoff $W$, impose $H_Xx=2a$ and $\sum_j x_j\le W$.
Let the rows of $L$ represent a basis of $\ker H_Z/\row H_X$.
Impose $Lx=2b+y$ with $y$ binary and $\sum_i y_i\ge1$.
The integer variables $a,b$ encode the parities, so a feasible support has zero syndrome and a nonzero logical class.
For each block $B_i$ of period $P$, impose $\sum_{q\in B_i}x_q\le P\sum_{j=0}^i x_{jP}$, indexing blocks from zero.
The first occupied block must then contain its origin.
This constraint covers the same $25$ starting coordinates as \codepkg{}, and a verified sector exchange transfers each exclusion to $X$.
The two integer solvers read the same model file.

On one Apple M3 Pro core with a $300$-second search limit, \codepkg{} excludes weights through $19$ for the $[[600,32]]$ input in $4.26$ seconds and through $27$ for the $[[2075,87]]$ input in $175.1$ seconds.
These runs use the proposed-addition test, with model preparation and input validation preceding timing.
HiGHS reaches the time limit on both inputs. Gurobi reaches the limit on the smaller input, while the available license prevents the larger run.

\label{sec:survey-case-details}
For the $2075$-qubit input \texttt{NM-P83}, we certify $d=34$.
Let $u_i$ be the determinant of the $3\times3$ minor obtained by deleting column $i$ from the published constituent $A(t)$ over $\mathbb F_2[t]/(t^{83}-1)$~\cite{NouroziEtAl2026Polynomial}.
Embedding $u$ in the first column of the $4\times4$ tensor block, with zeros elsewhere, gives a zero-syndrome $Z$ operator because $Au=0$.
The minor weights $8,8,10,8$ sum to $34$.
Sector exchange followed by a one-site translation gives an $X$ operator meeting its support in one qubit, proving both operators logical.
The LP reduction of Sec.~\ref{sec:lp-reduction} leaves nine full-code and four constituent starting coordinates.
The completed exclusions through weight $33$ and the verified sector exchange establish \mbox{$d_X=d_Z=34$}.
The $64$-worker calculation took approximately $14.3$ minutes of elapsed time on the reference scale, including preparation and the preceding lower-cutoff searches.

\subsection{Matched bound configurations and cutoff sweeps}
\label{app:mechanism-cutoffs}
One executable implements the four bound configurations in Table~\ref{tab:four-configurations} on twelve LP and two PP sector problems.
Cyclic starting coordinates, parity reduction, stabilizer pruning, the disjoint-neighborhood bound, and the order of candidate qubits remain fixed.
Every configuration selects an unsatisfied check with the fewest available neighbors, with overlap tie-breaking disabled.
Each of the fourteen problems receives three runs per configuration and a $120$-second search budget per run.
The runs follow a fixed shuffled order, one at a time, on an Apple M3 Pro.

\begin{table*}[tbp]
\centering\small\setlength{\tabcolsep}{3.7pt}
\begin{tabular}{@{}rrrccrrrr@{}}
\toprule
$(n,k)$ & $P$ & $w$ & Sector & $W$ & Off & \shortstack{Unsatisfied\\only} & \shortstack{Unsatisfied\\+ satisfied} & \shortstack{+ Proposed\\additions}\\
\midrule
$(1768,224)$~\cite{WebsterJacobHiggott2026Distance} & 52 & 8 & $Z$ & 21 & 23.6 & 26.4 & 6.97 & 7.04\\
$(1088,128)$~\cite{Lee2026RegularFrames} & 16 & 10 & $Z$ & 17 & 9.21 & 9.81 & 2.98 & 2.98\\
$(4350,1200)$~\cite{Lee2026Spectroscopy} & 75 & 19 & $Z$ & 14 & 34.1 & 7.73 & 1.62 & 1.56\\
$(5278,1456)$~\cite{Lee2026Spectroscopy} & 91 & 19 & $Z$ & 14 & 10.7 & 3.51 & 0.692 & 0.676\\
$(4770,784)$~\cite{Lee2026Spectroscopy} & 45 & 14 & $Z$ & 16 & 3.84 & 3.04 & 0.609 & 0.595\\
$(198,18)$~\cite{GuEtAl2026QGPU} & 11 & 8 & $X$ & 9 & 0.00238 & 0.00211 & 0.00129 & 0.00129\\
$(198,18)$~\cite{GuEtAl2026QGPU} & 11 & 8 & $Z$ & 9 & 0.00217 & 0.00198 & 0.00118 & 0.0012\\
$(442,68)$~\cite{WebsterJacobHiggott2026Distance} & 13 & 8 & $Z$ & 9 & 0.00067 & 0.000668 & 0.000489 & 0.000457\\
$(4075,175)$~\cite{KuoOuyang2026Erasure} & 163 & 7 & $Z$ & 21 & 6.8 & 7.58 & 2.51 & 2.46\\
$(4114,500)$~\cite{KuoOuyang2026Erasure} & 121 & 8 & $Z$ & 21 & 10.1 & 11 & 2.82 & 2.8\\
$(425,21)$~\cite{NouroziEtAl2026Polynomial} & 17 & 10 & $Z$ & 17 & 4.03 & 4.06 & 1.31 & 1.3\\
$(2075,87)$~\cite{NouroziEtAl2026Polynomial} & 83 & 10 & $Z$ & 21 & 5.43 & 6.28 & 1.19 & 1.18\\
\midrule
\multicolumn{9}{l}{Pair-partition codes~\cite{LeeEtAl2026DesignPrinciples}}\\
$(584,150)$ & 73 & 8 & $Z$ & 17 & 2.62 & 2.82 & 1.17 & 1.12\\
$(1112,282)$ & 139 & 8 & $Z$ & 19 & 10.6 & 11.4 & 5.14 & 4.99\\
\bottomrule
\end{tabular}
\caption{Successive additions to the weighted-check bound. Entries are median search times in seconds over three runs of one executable on an Apple M3 Pro. From left to right, the four timing columns disable the bound, use only unsatisfied checks, include satisfied checks, and also test proposed qubit additions. The branching and other pruning rules are fixed. All $168$ runs complete their exclusions. Search time includes bound evaluation and excludes preparation and initialization.}
\label{tab:four-configurations}
\end{table*}

\Needspace{3\baselineskip}
Testing proposed additions reduces node counts, with ratios of median times without and with this test ranging from $0.98$ to $1.07$ across the fourteen problems.

The cutoff sweeps use the $[[425,21,18]]$ and $[[1122,148,20]]$ codes.
The tested sector kernel of the $1122$-qubit code contains only even-weight words, while that of the $425$-qubit code also contains odd-weight words.
The requested cutoffs are the odd integers from $3$ through $17$ and $19$, respectively.
Each method receives the same physical input, test for logical nontriviality, cutoff after parity reduction, and cyclic restriction.
For the MaxCDCL model, the restriction is expressed by a hard clause
\begin{equation}
 x_v\ \Longrightarrow\ \bigvee_{j=0}^{i}x_{jP},\qquad v\in B_i,
\end{equation}
where $B_i$ is block $i$ of length $P$.
Thus the first occupied block contains its origin, and a common cyclic translation supplies such a representative for every nonempty support.
The mixed-integer models use the equivalent block-occupancy inequalities given above.
The connected-cluster search uses one launch for each required starting coordinate, with earlier coordinates forbidden.

The four \codepkg{} configurations and four baselines receive three serial repetitions at each cutoff, in fixed shuffled order, with a common $30$-second budget per run.
The timing boundary is total elapsed time as defined in Appendix~\ref{app:timing-conventions}.
Each run starts afresh, without reusing a lower bound from a preceding cutoff.
The MaxCDCL sweeps use \texttt{-no-elim} after the default returned an assignment violating a hard clause on a six-qubit cyclic control.
With variable elimination disabled, the cyclic model passes comparisons with exhaustive support enumeration.

\begin{table*}[tbp]
\centering\small
\begin{tabular}{@{}rlrrrr@{}}
\toprule
$n$ & Baseline & $W$ & $t_{\rm baseline}$ & $t_{\text{\codepkg}}$ & $t_{\rm baseline}/t_{\text{\codepkg}}$\\
\midrule
425 & m4riCC v0.9.0 & 9 & 1.5 & 0.0491 & 30.6\\
425 & MaxCDCL MSE 2023 & 5 & 14 & 0.0526 & 266\\
425 & Gurobi 13.0.3 & 9 & 4.56 & 0.0491 & 92.9\\
425 & HiGHS 1.15.1 & 7 & 6.7 & 0.0617 & 108\\
\midrule
1122 & m4riCC v0.9.0 & 11 & 6.62 & 0.0949 & 69.7\\
1122 & MaxCDCL MSE 2023 & 3 & 0.952 & 0.0827 & 11.5\\
1122 & Gurobi 13.0.3 & 9 & 2.95 & 0.0837 & 35.3\\
1122 & HiGHS 1.15.1 & 7 & 23.5 & 0.0862 & 273\\
\bottomrule
\end{tabular}
\caption{Largest tested cutoff $W$ at which the baseline and \codepkg{} complete all three repetitions within the common $30$-second limit. The $425$- and $1122$-qubit inputs are from Refs.~\cite{NouroziEtAl2026Polynomial} and~\cite{ZhouEtAl2026Surgery}, respectively. Entries give median elapsed times in seconds, including preparation. \codepkg{} uses constraints from unsatisfied and satisfied checks, with the proposed-addition test disabled. MaxCDCL uses \texttt{-no-elim}. Figure~\ref{fig:search-profile}(d) shows the $1122$-qubit sweep, including unfinished searches.}
\label{tab:shared-cutoffs}
\end{table*}

\Needspace{8\baselineskip}
\subsection{LP tensor-block comparison}
\label{app:lp-benchmark}
The nineteen monomial LP inputs in Ref.~\cite{WebsterJacobHiggott2026Distance} are tested in the $Z$ sector with matched general CSS and LP searches.
Each comparison runs on one cluster core.
The physical matrices, cutoff, internal branching rules, and pruning settings agree across configurations.
Each exclusion receives a budget of approximately $285$ seconds, shared by both parts of the LP search.
For the eleven inputs with $d_Z\ge20$, two repetitions reverse the configuration order and Table~\ref{tab:lp-specialization} reports their medians.
The remaining eight inputs have one repetition per configuration.
The known witnesses are checked on the physical matrices, and a verified even-weight constraint makes exclusion through $d_Z-2$ sufficient to establish the quoted sector distance.

Elapsed time includes program startup, matrix checks, construction of $S_{\cap}$, and both component searches.
Compilation, preparation of the common input files, and construction of complete logical detectors precede timing.
The implementation computes $S_{\cap}$ as the common kernel of the constituent checks and all restricted detectors $LE_j$.

\subsection{Parallel execution and runtime estimation}
\label{app:calibration}
The parallel comparison uses the same $16$ shards and \codepkg{} settings with one or four cluster workers. All three paired runs certify $d=22$ for the $1768$-qubit code with identical per-shard counts. Four workers give a median elapsed-time speedup of $3.31$ over one worker, including verification of the completed records.

For a witness-free exclusion, we estimate total search time by multiplying the mean time from a sample of starting coordinates by the number of required starting coordinates.
Across six fully measured sector problems, $10\,000$ samples of eight starting coordinates, drawn uniformly without replacement, gave estimates within a factor of two in $56.9\%$--$88.0\%$ of trials.

\subsection{Gross-code circuit benchmark}
\label{app:bb-circuit}
The circuit and noise model are taken from the software accompanying Ref.~\cite{BravyiEtAl2024Bicycle}, using its \href{https://github.com/sbravyi/BivariateBicycleCodes/blob/fa77e3333d3ec44c79d8f914dd24c040d1da471b/circuit_distance.py}{single-cycle distance experiment}.
The seven-layer CNOT schedule is followed for one noisy cycle and two noiseless cycles.
The circuit-level depolarizing model permits faults at CNOT gates, idle locations, initializations, and measurements, with each faulty operation costing one regardless of the number of qubits it affects.
All twelve logical qubits are monitored.
There are $3024$ enumerated single-fault outcomes per sector.
Merging identical detector and logical signatures and removing the zero signature leaves $936$ $X$-sector and $864$ $Z$-sector columns, each with $216$ detector rows and $12$ logical rows.
Verified lattice translations reduce the required starting coordinates to $13$ and $12$, respectively.
For this CSS extraction circuit, a nontrivial mixed-Pauli fault pattern has a nontrivial sector projection with no greater fault weight, so the two sector exclusions establish the full fault-distance lower bound.
The quoted search times are medians of three single-core runs on an Apple M3 Pro, with detector-model construction preceding timing.
We verify both ten-fault witnesses in the detector model and by replay through the source circuit.

\section{Catalogue of benchmark instances}
\label{app:literature-catalogue}
Tables~\ref{tab:survey-all} and~\ref{tab:bb-tt-certification} list the $84$ Abelian LP codes, and Table~\ref{tab:mitten-certification} lists the $11$ non-Abelian LP codes summarized in Table~\ref{tab:survey-family-summary}. We certify exact distances for $77$ and $11$ of these codes, respectively.
For two additional tricycle codes, Table~\ref{tab:bb-tt-certification} reports the $X$-sector distance only.
Tables~\ref{tab:survey-secondary} and~\ref{tab:pp-certification} give secondary Abelian LP inputs and six PP codes from Ref.~\cite{LeeEtAl2026DesignPrinciples}, respectively.

In the catalogue tables, $P$ denotes the verified cyclic period, $w$ the maximum binary check weight, and $w_c$ the maximum column weight.
Source status is U for an upper bound or estimate, E for an exact distance, ? for unclear certification, and N for no reported distance.
Bold entries denote exact distances certified here. Intervals combine our certified lower bounds with upper bounds from the cited sources or our searches.
\emph{Total search time} sums worker search times for the completed exclusions establishing the lower bound, on the Apple M3 Pro reference scale (Appendix~\ref{app:timing-conventions}).

\subsection{Sources and survey coverage}
\label{app:survey-sources}
We identify duplicate inputs by comparing binary stabilizer row spaces up to cyclic-block reversal and sector exchange.
The logical spectroscopy study includes the three Cain inputs and improves the $5278$-qubit upper bound to $22$. We count these inputs once.
The spectroscopy and Cain inputs with $(n,k)=(4350,1224)$ have distinct physical matrices and distances $16$ and $20$, respectively~\cite{Lee2026Spectroscopy,CainEtAl2026Shor}.

The survey includes the two principal radial examples and eight of the repository's $3000$ matrix pairs, all eleven polynomial-domain examples, and all sixteen cyclic constituents in Table~I of logical spectroscopy.
The $19$ Webster et al.\ inputs in Sec.~\ref{sec:lp-benchmark} and the $18$ monomial entries in Table~\ref{tab:survey-all} share $14$ parameter pairs $(n,k)$. The Webster benchmark includes five further lifts, and the catalogue includes four further LP04 codes.
We tested three original LP inputs of Ref.~\cite{SaboEtAl2024WeightReduction}. The permutations needed to reconstruct its twelve selected weight-reduced outputs were unavailable.

The eight non-Abelian inputs from Ref.~\cite{BhardwajEtAl2026Processors} comprise five reported distance values and three upper bounds.
The three inputs from Ref.~\cite{Hong2026NonAbelianLP} are reported with upper bounds.

For the BB and TT checks, we reconstruct the published polynomials over $C_\ell\times C_m$~\cite[Table~3]{BravyiEtAl2024Bicycle} and $C_\ell\times C_m\times C_p$~\cite[Tables~I and~V]{JacobMcLauchlanBrowne2025Tricycle}. Sector exchange suffices to transfer the BB exclusions. We search both sectors for TT code distances.

\Needspace{25\baselineskip}
\subsection{Catalogue runs and timing}
\label{sec:survey-timings}
\begin{wrapfigure}[19]{l}{0.391\textwidth}
\centering
\includegraphics[width=\linewidth]{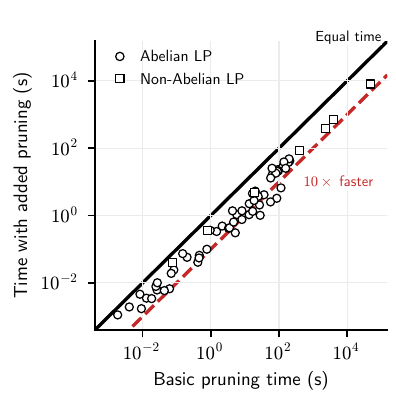}
\captionsetup{font={footnotesize,sf}}
\caption{Additional pruning reduces search time for $63$ matched catalogue exclusions.
Both axes use the Apple M3 Pro reference scale.
Black solid and red dashed lines indicate equal times and a tenfold speedup.}
\label{fig:catalogue-refresh}
\end{wrapfigure}

Figure~\ref{fig:catalogue-refresh} compares $63$ catalogue exclusions with matched inputs, cutoffs, symmetries, and shard settings.
Both searches use symmetry reduction, stabilizer pruning, and the disjoint-neighborhood bound.
Adding weighted-check and proposed-addition pruning with overlap-based check selection reduces search time in every case, with a median speedup of $4.94\times$.
Figure~\ref{fig:search-profile}(c) isolates the weighted-check bound.
The exclusions in Fig.~\ref{fig:catalogue-effort} use the weighted-check bound and have $n\ge144$, a completed lower bound of at least ten, and a recorded processor model.
Of the $58$ Abelian LP codes for which we certified a published upper bound, $43$ took under five seconds of combined search time on the Apple M3 Pro reference scale, including the distance-$20$ code in Appendix~\ref{app:polynomial-comparison}.

In Fig.~\ref{fig:catalogue-refresh}, times with additional pruning are medians of three complete exclusions. Short exclusions were timed directly on one Apple M3 Pro core, and the remaining repeats used cluster cores. The $170$-qubit PP and $756$-qubit BB runs also used the cluster, with up to $32$ workers for PP. The three Hong codes used $32$ workers each.

\label{app:conditional-followup}
The high-check-weight follow-up covers nine initially unresolved Abelian LP inputs and fourteen exact-distance controls, all with $w\ge10$ and verified sector exchange. Five of the nine lower bounds improve, and two exclusions establish exact distances. Each code has a $300$-second budget across cutoffs and sectors on one M3 Pro worker.
Overlap tie-breaking and the proposed-addition test are enabled. Parity reductions use a verified row certificate. All fourteen controls complete through $d-1$.

\begingroup
\footnotesize
\setlength{\tabcolsep}{3pt}
\setlength{\LTcapwidth}{\textwidth}
\begin{longtable}{@{}llrrclcr@{}}
\caption{Abelian matrix-LP survey. For \texttt{RA-4-23N} and \texttt{NM-P83}, the first rows give exclusions at lower cutoffs. Starred rows complete the exact-distance certification, using general CSS search for \texttt{RA-4-23N} and the LP reduction of Sec.~\ref{sec:lp-reduction} for \texttt{NM-P83}.}\label{tab:survey-all}\\
\toprule
ID & $(n,k)$ & $P$ & $w$ & Source $d$ & Status & This work & Total search time (s) \\
\midrule
\endfirsthead
\caption[]{Abelian matrix-LP survey (continued).}\\
\toprule
ID & $(n,k)$ & $P$ & $w$ & Source $d$ & Status & This work & Total search time (s) \\
\midrule
\endhead
\bottomrule
\endfoot
\bottomrule
\endlastfoot
\multicolumn{8}{l}{\textit{Monomial LP benchmarks}}\\*
\texttt{LP04-175}~\cite{RaveendranEtAl2022GKP} & $(175,19)$ & $7$ & $7$ & $\le 10$ & U & $\mathbf{10}$ & $<0.01$ \\
\texttt{LP04-225} & $(225,21)$ & $9$ & $7$ & $\le 12$ & U & $\mathbf{12}$ & $<0.01$ \\
\texttt{LP04-425} & $(425,29)$ & $17$ & $7$ & $\le 18$ & U & $\mathbf{18}$ & $0.487$ \\
\texttt{LP04-475} & $(475,31)$ & $19$ & $7$ & $\le 20$ & U & $\mathbf{20}$ & $2.25$ \\
\texttt{LP118-544}~\cite{RaveendranEtAl2022GKP} & $(544,80)$ & $16$ & $8$ & $12$~\cite{WebsterJacobHiggott2026Distance} & E & $\mathbf{12}$ & $<0.01$ \\
\texttt{LP118-714} & $(714,100)$ & $21$ & $8$ & $\le 16$ & U & $\mathbf{16}$ & $0.1$ \\
\texttt{LP118-1020} & $(1020,136)$ & $30$ & $8$ & $\le 16$~\cite{WebsterJacobHiggott2026Distance} & U & $\mathbf{16}$ & $0.0658$ \\
\texttt{Tanner-1054}~\cite{PanteleevKalachev2022} & $(1054,140)$ & $31$ & $8$ & $\le 20$ & U & $\mathbf{20}$ & $2.07$ \\
\texttt{Xu-1428}~\cite{XuEtAl2024AtomArrays} & $(1428,184)$ & $42$ & $8$ & $\le 24$ & U & $\mathbf{24}$ & $48.3$ \\
\texttt{WG-AJ52}~\cite{WebsterJacobHiggott2026Distance} & $(1768,224)$ & $52$ & $8$ & $\le 22$ & U & $\mathbf{22}$ & $6.64$ \\
\texttt{WG-AJ58} & $(1972,248)$ & $58$ & $8$ & $\le 20$ & U & $\mathbf{20}$ & $1.06$ \\
\texttt{WG-AJ60} & $(2040,256)$ & $60$ & $8$ & $\le 20$ & U & $\mathbf{20}$ & $1.34$ \\
\texttt{WG-AJ61} & $(2074,260)$ & $61$ & $8$ & $\le 24$ & U & $\mathbf{24}$ & $28.2$ \\
\texttt{WG-AJ67} & $(2278,284)$ & $67$ & $8$ & $\le 22$ & U & $\mathbf{22}$ & $5.33$ \\
\texttt{WG-AJ73} & $(2482,308)$ & $73$ & $8$ & $\le 24$ & U & $\mathbf{24}$ & $21.5$ \\
\texttt{WG-AJ80} & $(2720,336)$ & $80$ & $8$ & $\le 22$ & U & $\mathbf{22}$ & $4.53$ \\
\texttt{WG-AJ85} & $(2890,356)$ & $85$ & $8$ & $\le 24$ & U & $\mathbf{24}$ & $20.5$ \\
\texttt{WG-AJ95} & $(3230,396)$ & $95$ & $8$ & $\le 24$ & U & $\mathbf{24}$ & $18.2$ \\
\addlinespace
\multicolumn{8}{l}{\textit{Logical spectroscopy}~\cite{Lee2026Spectroscopy}}\\*
\texttt{LS-P15-2X4} & $(300,60)$ & $15$ & $11$ & $9$ & E & $\mathbf{9}$ & $<0.01$ \\
\texttt{LS-E14-3X4} & $(350,26)$ & $14$ & $7$ & $\le 14$ & U & $\mathbf{14}$ & $0.0242$ \\
\texttt{LS-R21-4X5} & $(861,45)$ & $21$ & $9$ & $\le 26$ & U & $[23,26]$ & $25.3$ \\
\texttt{LS-P35-3X4} & $(875,35)$ & $35$ & $13$ & $\le 35$ & U & $[21,35]$ & $151$ \\
\texttt{LS-P31-2X5} & $(899,279)$ & $31$ & $13$ & $\le 8$ & U & $\mathbf{8}$ & $<0.01$ \\
\texttt{LS-H33-3X9} & $(2970,1188)$ & $33$ & $23$ & $\le 18$ & U & $[13,18]$ & $161$ \\
\texttt{LS-H41-3X8} & $(2993,1025)$ & $41$ & $21$ & $\le 26$ & U & $[14,26]$ & $53.3$ \\
\texttt{LS-P75-3X7} & $(4350,1200)$ & $75$ & $19$ & $\le 25$ & U & $[19,25]$ & $420$ \\
\texttt{LS-R75-3X7} & $(4350,1224)$ & $75$ & $10$ & $\le 16$ & U & $\mathbf{16}$ & $0.433$ \\
\texttt{LS-R81-3X7} & $(4698,1320)$ & $81$ & $10$ & $\le 16$ & U & $\mathbf{16}$ & $0.642$ \\
\texttt{LS-M45-5X9} & $(4770,784)$ & $45$ & $14$ & $\le 50$ & U & $[22,50]$ & $40.9$ \\
\texttt{LS-M45-4X10} & $(5220,1674)$ & $45$ & $14$ & $\le 16$ & U & $\mathbf{16}$ & $2.55$ \\
\texttt{LS-P91-3X7} & $(5278,1456)$ & $91$ & $19$ & $\le 70$ & U & $[19,70]$ & $117$ \\
\addlinespace
\multicolumn{8}{l}{\textit{High-rate monomial LP inputs}~\cite{CainEtAl2026Shor}}\\*
\texttt{Cain-2610} & $(2610,744)$ & $45$ & $10$ & $\le 16$ & U & $\mathbf{16}$ & $0.763$ \\
\texttt{Cain-4350} & $(4350,1224)$ & $75$ & $10$ & $\le 20$ & U & $\mathbf{20}$ & $25.4$ \\
\texttt{Cain-5278} & $(5278,1480)$ & $91$ & $10$ & $\le 22$~\cite{Lee2026Spectroscopy} & U & $\mathbf{22}$ & $164$ \\
\addlinespace
\multicolumn{8}{l}{\textit{Radial codes}~\cite{ScrubyHillmannRoffe2026Radial}}\\*
\texttt{Radial-90} & $(90,8)$ & $5$ & $6$ & $\le 10$ & U & $\mathbf{10}$ & $<0.01$ \\
\texttt{RA-3-7D} & $(126,14)$ & $7$ & $6$ & $\le 8$ & U & $\mathbf{8}$ & $<0.01$ \\
\texttt{RA-4-7N} & $(224,18)$ & $7$ & $8$ & $\le 14$ & U & $\mathbf{14}$ & $0.0411$ \\
\texttt{RA-3-13D} & $(234,8)$ & $13$ & $6$ & $\le 12$ & U & $\mathbf{12}$ & $<0.01$ \\
\texttt{RA-3-13N} & $(234,8)$ & $13$ & $6$ & $\le 14$ & U & $\mathbf{14}$ & $<0.01$ \\
\texttt{RA-3-17D} & $(306,8)$ & $17$ & $6$ & $\le 14$ & U & $\mathbf{14}$ & $<0.01$ \\
\texttt{Radial-352} & $(352,18)$ & $11$ & $8$ & $\le 20$ & U & $\mathbf{20}$ & $3.26$ \\
\texttt{RA-4-13N} & $(416,18)$ & $13$ & $8$ & $\le 18$ & U & $\mathbf{18}$ & $0.308$ \\
\texttt{RA-4-17N} & $(544,18)$ & $17$ & $8$ & $\le 22$ & U & $\mathbf{22}$ & $4.15$ \\
\texttt{RA-4-23N} & $(736,18)$ & $23$ & $8$ & $\le 26$ & U & $[24,26]$ & $9.34$ \\*
\texttt{RA-4-23N}$^{*}$ & & & & & & $\mathbf{26}$ & $64.9$ \\
\addlinespace
\multicolumn{8}{l}{\textit{Weight-reduction inputs}~\cite{SaboEtAl2024WeightReduction}}\\*
\texttt{Sabo-260} & $(260,58)$ & $13$ & $6$ & $\le 6$ & U & $\mathbf{6}$ & $<0.01$ \\
\texttt{Sabo-775} & $(775,43)$ & $31$ & $7$ & $\le 24$ & U & $\mathbf{24}$ & $37.6$ \\
\texttt{Sabo-1150} & $(1150,50)$ & $46$ & $7$ & $\le 21$ & U & $\mathbf{21}$ & $2.8$ \\
\addlinespace
\multicolumn{8}{l}{\textit{Bias-tailoring CSS input}~\cite{RoffeEtAl2023BiasTailored}}\\*
\texttt{Roffe-416} & $(416,18)$ & $13$ & $8$ & $\le 20$ & U & $\mathbf{20}$ & $1.02$ \\
\addlinespace
\multicolumn{8}{l}{\textit{Polynomial-domain LP codes}~\cite{NouroziEtAl2026Polynomial}}\\*
\texttt{NM-P10} & $(250,18)$ & $10$ & $10$ & $8$ & ? & $\mathbf{8}$ & $<0.01$ \\
\texttt{NM-P11} & $(275,15)$ & $11$ & $8$ & $12$ & ? & $\mathbf{12}$ & $<0.01$ \\
\texttt{NM-P12} & $(300,20)$ & $12$ & $10$ & $12$ & ? & $\mathbf{12}$ & $0.0101$ \\
\texttt{NM-P13} & $(325,17)$ & $13$ & $10$ & $14$ & ? & $\mathbf{14}$ & $0.0576$ \\
\texttt{NM-P14} & $(350,22)$ & $14$ & $8$ & $12$ & ? & $\mathbf{12}$ & $<0.01$ \\
\texttt{NM-P15} & $(375,19)$ & $15$ & $10$ & $16$ & ? & $\mathbf{16}$ & $0.337$ \\
\texttt{NM-P16} & $(400,24)$ & $16$ & $10$ & $13$ & ? & $\mathbf{13}$ & $0.0192$ \\
\texttt{NM-P17} & $(425,21)$ & $17$ & $10$ & $18$ & ? & $\mathbf{18}$ & $1.08$ \\
\texttt{NM-P24} & $(600,32)$ & $24$ & $10$ & $\le 20$ & U & $\mathbf{20}$ & $3.71$ \\
\texttt{NM-P41} & $(1025,45)$ & $41$ & $10$ & $\le 26$ & U & $\mathbf{26}$ & $152$ \\
\texttt{NM-P83} & $(2075,87)$ & $83$ & $10$ & $\le 34$ & U & $[28,34]$ & $195$ \\*
\texttt{NM-P83}$^{*}$ & & & & & & $\mathbf{34}$ & $20500$ \\
\addlinespace
\multicolumn{8}{l}{\textit{Clustered-cyclic LP codes}~\cite{GuEtAl2026QGPU}}\\*
\texttt{QGPU-8-7} & $(56,8)$ & $7$ & $8$ & $\le 7$ & U & $\mathbf{7}$ & $<0.01$ \\
\texttt{QGPU-8-11} & $(88,8)$ & $11$ & $8$ & $\le 10$ & U & $\mathbf{10}$ & $0.0114$ \\
\texttt{QGPU-8-13} & $(104,8)$ & $13$ & $8$ & $\le 11$ & U & $\mathbf{11}$ & $0.0197$ \\
\texttt{QGPU-18-7} & $(126,18)$ & $7$ & $8$ & $\le 7$ & U & $\mathbf{7}$ & $<0.01$ \\
\texttt{QGPU-8-17} & $(136,8)$ & $17$ & $8$ & $\le 14$ & U & $\mathbf{14}$ & $0.4$ \\
\texttt{QGPU-18-11} & $(198,18)$ & $11$ & $8$ & $\le 10$ & U & $\mathbf{10}$ & $<0.01$ \\
\addlinespace
\multicolumn{8}{l}{\textit{Erasure-decoding LP inputs}~\cite{KuoOuyang2026Erasure}}\\*
\texttt{KO-925} & $(925,49)$ & $37$ & $7$ & --- & N & $\mathbf{20}$ & $1.38$ \\
\texttt{KO-2075} & $(2075,95)$ & $83$ & $7$ & --- & N & $\mathbf{24}$ & $16.9$ \\
\texttt{KO-2210} & $(2210,276)$ & $65$ & $8$ & --- & N & $\mathbf{24}$ & $39.1$ \\
\texttt{KO-4075} & $(4075,175)$ & $163$ & $7$ & --- & N & $\mathbf{24}$ & $11.1$ \\
\texttt{KO-4114} & $(4114,500)$ & $121$ & $8$ & --- & N & $\mathbf{24}$ & $15.7$ \\
\addlinespace
\multicolumn{8}{l}{\textit{Finite-length distance examples}~\cite{RaveendranDeclercqVasic2025}}\\*
\texttt{RDV600-1} & $(600,36)$ & $24$ & $7$ & $7$ & ? & $\mathbf{7}$ & $<0.01$ \\
\texttt{RDV600-2} & $(600,36)$ & $24$ & $7$ & $20$ & ? & $\mathbf{20}$ & $1.39$ \\
\addlinespace
\multicolumn{8}{l}{\textit{Other Abelian matrix LP examples}}\\*
\texttt{LCS-369}~\cite{OldRisplerMueller2024LCS} & $(369,9)$ & $9$ & $6$ & $9$ & ? & $\mathbf{9}$ & $<0.01$ \\
\texttt{Cyclic-560}~\cite{BhardwajEtAl2026Processors} & $(560,112)$ & $28$ & $9$ & $\le 14$ & U & $\mathbf{14}$ & $0.366$ \\
\addlinespace
\multicolumn{8}{l}{\textit{Cyclic spatially coupled example}~\cite{LiuMarquardt2026}}\\*
\texttt{SCE-R1E01} & $(1428,186)$ & $42$ & $8$ & $\le 18$ & U & $\mathbf{18}$ & $0.42$ \\
\end{longtable}
\endgroup

\normalsize
\begin{center}
\begin{minipage}{\textwidth}
\centering\footnotesize\setlength{\tabcolsep}{5pt}
\captionof{table}{Non-Abelian LP codes, all with $w=9$. The Mitten and Hong rows use the constructions of Refs.~\cite{BhardwajEtAl2026Processors} and~\cite{Hong2026NonAbelianLP}, respectively.
$G$ is the construction group, and $D_{10}$ has order ten.
The starred row repeats the $975$-qubit certification using the LP reduction of Sec.~\ref{sec:lp-reduction}, with its time measured in one complete execution.}\label{tab:mitten-certification}
\begin{tabular}{@{}lllrccr@{}}
\toprule
ID & $(n,k)$ & $G$ & $P$ & Source $d$ & Certified $d$ & Total search time (s)\\
\midrule
\texttt{Mitten-150} & $(150,30)$ & $C_5\times S_3$ & 5 & 10 & $\mathbf{10}$ & $0.0395$ \\
\texttt{Mitten-200} & $(200,40)$ & $C_4\times D_{10}$ & 4 & 12 & $\mathbf{12}$ & $0.36$ \\
\texttt{Mitten-300} & $(300,60)$ & $C_{10}\times S_3$ & 10 & 14 & $\mathbf{14}$ & $0.894$ \\
\texttt{Hong-390} & $(390,78)$ & $C_{26}\rtimes C_3$ & 2 & $\le16$ & $\mathbf{16}$ & 25.4\\
\texttt{Mitten-500} & $(500,100)$ & $C_5\times(C_5\rtimes C_4)$ & 5 & 16 & $\mathbf{16}$ & $4.91$ \\
\texttt{Mitten-540} & $(540,108)$ & $(C_9\rtimes C_3)\rtimes C_4$ & 2 & 18 & $\mathbf{18}$ & $84.7$ \\
\texttt{Hong-550} & $(550,110)$ & $C_{22}\rtimes C_5$ & 2 & $\le18$ & $\mathbf{18}$ & 101.6\\
\texttt{Hong-625} & $(625,125)$ & $C_{25}\rtimes C_5$ & 5 & $\le20$ & $\mathbf{20}$ & 430.4\\
\texttt{Mitten-630} & $(630,126)$ & $C_7\rtimes C_{18}$ & 3 & $\le20$ & $\mathbf{20}$ & $391$ \\
\texttt{Mitten-780} & $(780,156)$ & $C_{13}\times A_4$ & 13 & $\le22$ & $\mathbf{22}$ & $714$ \\
\texttt{Mitten-975} & $(975,195)$ & $C_5\times(C_{13}\rtimes C_3)$ & 5 & $\le24$ & $\mathbf{24}$ & $8056$ \\
\texttt{Mitten-975}$^{*}$ & & & & & $\mathbf{24}$ & $2579$ \\
\bottomrule
\end{tabular}
\end{minipage}
\end{center}
\small
\begin{center}
\begin{minipage}{\textwidth}
\centering\footnotesize\setlength{\tabcolsep}{3pt}
\captionof{table}{Supplemental LP inputs. The scalar-by-matrix codes provide earlier benchmarks beyond the families selected for Table~\ref{tab:survey-family-summary}. Source-discrepancy cases document differences between printed matrices and reported distances, and diagnostic constituents serve as structural tests. These three supplemental groups are excluded from the principal-survey totals. I denotes a published rigorous interval, X an unresolved source discrepancy, and D a diagnostic constituent. Results for X rows apply to the printed matrices.}\label{tab:survey-secondary}
\begin{tabular}{@{}llrrclcr@{}}
\toprule
ID & $(n,k)$ & $P$ & $w$ & Source $d$ & Status & This work & Total search time (s) \\
\midrule
\multicolumn{8}{l}{\textit{Scalar-by-matrix LP codes}}\\
\texttt{PK-B1-882}~\cite{PanteleevKalachev2021} & $(882,24)$ & $63$ & $6$ & $[18,24]$ & I & $\mathbf{24}$ & $23.1$ \\
\texttt{KO-GHP882}~\cite{KuoOuyang2026Erasure,KuoLai2022Degeneracy} & $(882,48)$ & $63$ & $8$ & $16$ & ? & $\mathbf{16}$ & $0.0741$ \\
\texttt{PK-B3-1270}~\cite{PanteleevKalachev2021} & $(1270,28)$ & $127$ & $6$ & $[16,46]$ & I & $[24,46]$ & $13.1$ \\
\addlinespace
\multicolumn{8}{l}{\textit{Printed matrices}~\cite{PradhanEtAl2026FastDecoders}}\\
\texttt{Pradhan-printed-775} & $(775,43)$ & $31$ & $7$ & $\le 20$ & X & $\mathbf{24}$ & $41.9$ \\
\texttt{Pradhan-printed-1054} & $(1054,140)$ & $31$ & $8$ & $\le 20$ & X & $\mathbf{16}$ & $0.0552$ \\
\addlinespace
\multicolumn{8}{l}{\textit{Spectroscopy diagnostic constituents}~\cite{Lee2026Spectroscopy}}\\
\texttt{LS-RES7-3X4} & $(175,19)$ & $7$ & $7$ & --- & D & $\mathbf{10}$ & $<0.01$ \\
\texttt{LS-CMP15-2X4} & $(300,66)$ & $15$ & $6$ & $6$ & D & $\mathbf{6}$ & $<0.01$ \\
\addlinespace
\bottomrule
\end{tabular}
\end{minipage}
\end{center}
\normalsize
\begin{center}
\begin{minipage}{\textwidth}
\centering\small\setlength{\tabcolsep}{5pt}
\captionof{table}{PP distance benchmarks for six codes certified in Ref.~\cite{LeeEtAl2026DesignPrinciples}.}\label{tab:pp-certification}
\begin{tabular}{@{}rrrrrr@{}}
\toprule
$(n,k)$ & $P$ & $w$ & $w_c$ & $d$ & Total search time (s) \\
\midrule
$(170,8)$ & 85 & 10 & 5 & 20 & 2110 \\
$(248,66)$ & 31 & 8 & 3 & 12 & 0.00708 \\
$(372,130)$ & 31 & 12 & 4 & 18 & 158 \\
$(584,150)$ & 73 & 8 & 3 & 18 & 2.31 \\
$(1112,282)$ & 139 & 8 & 3 & 20 & 10.7 \\
$(1784,450)$ & 223 & 8 & 3 & 24 & 419 \\
\bottomrule
\end{tabular}
\end{minipage}
\end{center}

\begin{center}
\begin{minipage}{\textwidth}
\centering\small\setlength{\tabcolsep}{6pt}
\captionof{table}{Distance certification for BB codes~\cite{BravyiEtAl2024Bicycle} and TT codes~\cite{JacobMcLauchlanBrowne2025Tricycle}, using the general CSS search and full translation group $G$.
The first two groups give full-code distances, and the final group gives $X$-sector distances.
Local timings are medians of three runs.
The $756$-qubit BB exclusion used $256$ parallel partitions and took approximately $35.1$ minutes of parallel elapsed time.
All times use the Apple M3 Pro reference scale, and the table entries give total worker search time.}
\label{tab:bb-tt-certification}
\begin{tabular}{@{}rcrrcr@{}}
\toprule
$(n,k)$ & $G$ & $W$ & Source value & This work & Total search time (s) \\
\midrule
\multicolumn{6}{l}{BB codes, $d=d_X=d_Z$, $w=6$, $w_c=3$} \\
$(288,12)$ & $C_{12}^2$ & 17 & $18$ & $\mathbf{18}$ & 0.0675 \\
$(360,12)$ & $C_{30}\times C_6$ & 23 & $\le24$ & $\mathbf{24}$ & 36.6 \\
$(756,16)$ & $C_{21}\times C_{18}$ & 33 & $\le34$ & $\mathbf{34}$ & 123000 \\
\midrule
\multicolumn{6}{l}{TT codes, $d=d_Z$, $w=9$, $w_c=6$} \\
$(588,9)$ & $C_7^2\times C_4$ & 13 & $\le14$ & $\mathbf{14}$ & 0.0362 \\
$(648,6)$ & $C_6^3$ & 15 & $\le16$ & $\mathbf{16}$ & 0.510 \\
$(648,12)$ & $C_6^3$ & 13 & $\le14$ & $\mathbf{14}$ & 0.0290 \\
$(735,18)$ & $C_7^2\times C_5$ & 13 & $\le14$ & $\mathbf{14}$ & 0.0261 \\
$(840,9)$ & $C_8\times C_7\times C_5$ & 15 & $\le16$ & $\mathbf{16}$ & 0.245 \\
$(1029,18)$ & $C_7^3$ & 15 & $\le16$ & $\mathbf{16}$ & 0.156 \\
\midrule
\multicolumn{6}{l}{TT codes, $X$-sector distance $d_X$, $w=9$, $w_c=6$} \\
$(180,12)$ & $C_5\times C_4\times C_3$ & 19 & $\le20$ & $\mathbf{20}$ & 0.0132 \\
$(432,12)$ & $C_6^2\times C_4$ & 35 & $\le36$ & $\mathbf{36}$ & 25.9 \\
\bottomrule
\end{tabular}
\end{minipage}
\end{center}

\end{document}